\documentclass[preprint,aps,showpacs]{revtex4}
\usepackage{graphicx}
\usepackage{epsfig,subfigure}
\usepackage{amsmath}
 
\begin{document}
\newcommand{\bvec}[1]{\mbox{\boldmath ${#1}$}}
\title{Photo- and Electroproduction of the Hypertriton on $^3$He}
\author{T.\ Mart$^1$ and B.\ I.\ S.\ van der Ventel$^2$} 
\affiliation{$^1$Departemen Fisika, FMIPA, Universitas Indonesia, 
        Depok 16424, Indonesia\\
      $^2$Department of Physics, Stellenbosch University, Private Bag X1, 
    Matieland 7602, South Africa}
\date{\today}
\begin{abstract}
 Differential cross sections of the photo- and 
 electroproduction of the hypertriton have been calculated by utilizing
 modern nuclear wave functions and the elementary operator of KAON-MAID. 
 It is found that a proper treatment of Fermi motion is essential 
 for the two processes. While the average momentum approximation  
 can partly simulate the Fermi motion in the process, 
 the ``frozen nucleon'' assumption yields very different results, especially at
 lower energies. The Coulomb effect induced by the interaction between the
 positively charged kaon and the hypertriton 
 is found to be negligible. The influence of higher
 partial waves is also found to be relatively small, in contrast
 to the finding of the previous work. The off-shell
 assumption is found to be very sensitive in the
 case of electroproduction rather than in photoproduction.
 It is shown that the few available experimental
 data favor the assumption that the initial nucleon
 is off-shell and the final hyperon is on-shell.
 This seems to be reasonable, since the hyperon
 in the hypertriton is less bound than the nucleon
 in the initial $^3$He nucleus.  
 The effect of the  missing resonance $D_{13}(1895)$
 is more profound in the longitudinal cross sections. 
 Excluding this resonance reduces the longitudinal 
 cross sections by one order of magnitude, but does not
 change the effects of various off-shell 
 assumptions on the cross sections.
\end{abstract}
\pacs{13.60.Le, 25.20.Lj, 21.80.+a}

\maketitle


\section{Introduction}
The hypertriton  is a bound state consisting of a proton, a neutron, and 
a $\Lambda$ hyperon. Although a hypertriton consisting of a proton, 
a neutron, and a $\Sigma^0$ hyperon could exist, no experimental
information is available at present \cite{sighyp,Afnan,Dover}. Therefore, we will 
use the term ``hypertriton'' to denote the $\Lambda$-hypertriton in 
the following. Interest in the hypertriton is mainly due to the fact that  
it is the lightest and the loosely bound hypernucleus. The separation
energy into a $\Lambda$ and a deuteron is only $0.13\pm 0.05$ 
MeV~\cite{Juric:1973zq}, while the total binding energy is 2.35 MeV.
Being the lightest hypernucleus, the hypertriton
is obviously the first system in which the 
$YN$ potential, including the interesting $\Lambda$-$\Sigma$ conversion, 
can be tested in the nuclear medium. This is also supported by the fact
that neither the $\Lambda N$ nor the $\Sigma N$ interactions possesses 
sufficient strength to produce a bound two-body system, while on the other
hand the available $YN$ scattering data are still extremely poor. 
Therefore the hypertriton is expected to play an important role in 
hypernuclear physics similar to that of the deuteron in conventional nuclear physics.
Due to experimental difficulties, however, the existing information on the 
hypertriton is mostly from old measurements \cite{oldi}.

Recently, theoretical investigation of the hypertriton properties 
have drawn considerable attentions in the nuclear physics 
community~\cite{miyagawa93,miyagawa95,hyp_properties}. 
The Bochum group~\cite{miyagawa93} has investigated 
the hypertriton by using various $YN$ and $NN$ potentials. 
Interestingly, when the J\"ulich hyperon-nucleon 
potential in the one-boson-exchange (OBE) parameterization 
(model $\tilde{\rm A}$ of Ref.~\cite{juelich}) combined with 
various realistic $NN$ interactions were 
used, then the hypertriton turned out to be unbound. 
Only an increase by about 4\% in the J\"ulich potential can bring
the hypertriton back to a bound state. However, the use of the Nijmegen 
hyperon-nucleon potential in the same calculation~\cite{nijmegen89} 
leads to a bound hypertriton. This fact indicates that significant 
improvement in the hyperon-nucleon force sector is strongly needed.

In principle, the hypertriton could be produced by employing hadronic properties
such as stopped and low momentum kaon induced reactions,
A$(K,\pi)$B and A$(\pi,K)$B. 
Another possibility to obtain hypertriton is by utilizing the proton-deuteron
collision
\begin{eqnarray}
  p + d \to K^+ + {^3_\Lambda{\rm H}} ~.
\end{eqnarray}
Komarov {\it et al}. have studied this process theoretically at
incident proton energies $T_p=1.13-3.0$ GeV and found that the
cross section is on the order of 1 nb, at most~\cite{komarov}. This
result has been refined in Ref.~\cite{Gardestig:1996zz} by using a two-step model and
the differential cross section is found to be much smaller than 
1 nb/sr.

Because the electromagnetic beams (electrons or real photons) are 
well understood, clean, and well under control, the use of the 
electromagnetic processes has, however, a competing advantage
compared to the hadronic ones. 
More than one decade ago one of the authors has estimated the
cross section of the hypertriton photoproduction 
\begin{eqnarray}
  \gamma + {^3{\rm He}} \to K^+ + {^3_\Lambda{\rm H}} ~,
  \label{eq:hyp_photo}
\end{eqnarray}
and investigated the effects of the off-shell assumption and 
Fermi motion on this process~\cite{mart98,mart_thesis}.
This has been performed by using the 
wave function of $^3$He obtained as a 
solution of the Faddeev equations with the Reid soft core potential 
\cite{kim}, and a simple hypertriton wave function developed in Ref. 
\cite{congleton}, along with the elementary operator from
Williams {\it et al.}~\cite{williams}. The result showed that
the cross sections are predicted to be on the order of 1 nb and
drop quickly as a function of the kaon scattering angle. Eight
years later three experimental data points on the hypertriton
electroproduction
\begin{eqnarray}
  e + {^3{\rm He}} \to e' + K^+ + {^3_\Lambda{\rm H}} ~,
  \label{eq:hyp_electro}
\end{eqnarray}
at $\theta_K^{\rm c.m.}=2.7^\circ,9.5^\circ$, and $18.9^\circ$, 
were published by Dohrmann {\it et al.}~\cite{Dohrmann:2004xy}. 
Although this process utilizes electrons (virtual photons), 
and therefore is different from the process given by Eq.~(\ref{eq:hyp_photo}), 
the result shows a surprising phenomenon, i.e., the angular distribution of 
the differential cross section shows an almost flat structure.
An extrapolation of the photoproduction result to the finite $k^2$
region is only able to reproduce the trend of the first two data 
points and, on the other hand, underpredicts the last data point by one order of 
magnitude~\cite{note_on_dohrmann}. This is in contrast to the process
\begin{eqnarray}
  e + {^4{\rm He}} \to e' + K^+ + {^4_\Lambda{\rm H}} ~,
\end{eqnarray}
reported by the same experiment~\cite{Dohrmann:2004xy}, for which 
the cross section decreases smoothly and nicely fits the 
prediction~\cite{note_on_dohrmann1}.

The present work has been greatly motivated by the facts described above. 
In the present work we shall only focus on the photo- and 
electroproduction of the hypertriton and leave the electroproduction 
of $^4_\Lambda{\rm H}$ for the future consideration.
For this purpose we shall use the modern nuclear wave 
functions~\cite{nijmegen93,miyagawa93} as well as the frequently
used elementary operator KAON-MAID~\cite{kaon-maid}
to study the effects of the various off-shell assumptions, 
Fermi motion, and Coulomb interaction between the exited kaon
and the hypertriton, 
on the calculated differential cross sections. Recently, this
elementary operator has been used to investigate the
final states $YN$ and $KN$ interactions in kaon photoproduction
off a deuteron as well as to investigate the possibility
of extracting the elementary process $N(\gamma,K^0)Y$ from this process
at the quasi-free-scattering kinematics~\cite{YaM99}.
The elementary
operator is given in a unique form that is completely frame 
independent, since it can be expressed in terms of the Mandelstam
variables $s$, $u$, and $t$, or the four-momenta of the photon,
nucleon, kaon, and hyperon. Furthermore, the operator does not contain 
the photon-polarization-vector $\epsilon^\mu$ and spin-operator $\sigma^{(n)}$ 
terms. This guarantees the analytical continuation of the elementary
amplitude and enables us to use different off-shell assumptions. Thus,
the result would provide us with a refined calculation of the hypertriton
photoproduction and a direct comparison with the electroproduction
data.

This paper is organized as follows: In Section \ref{sec:elementary} 
we shall briefly review the properties of the elementary operator used in 
this work. Section \ref{sec:nuclear_op} presents the formalism of 
the nuclear operator along with its relation to the elementary
operator and to the nuclear cross sections. We shall present and discuss 
the results of our calculations in Section \ref{sec:result}. 
Section \ref{sec:conclusion} summarizes our findings. A few important notes 
on the elementary amplitudes, the anti-symmetry factor of the nuclear wave 
functions, and some kinematical relations are given in the Appendices.

\section{Properties of the Elementary Operator}\label{sec:elementary}
\begin{figure}[!t]
  \begin{center}
    \leavevmode
    \epsfig{figure=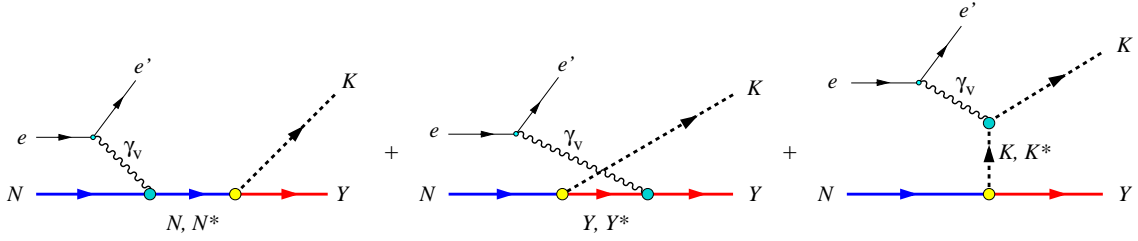,width=150mm}
    \caption{(Color online) The basic Feynman diagrams in the 
        elementary operator. }
   \label{fig:feynman} 
  \end{center}
\end{figure}

Since photoproduction is only a special case of electroproduction,
we will only consider the latter in our formalism. The results for
photoproduction are obtained by setting the virtual photon momentum
to zero.
The elementary process for electroproduction of a kaon and a hyperon
on the nucleon target can be written as
\begin{eqnarray}
  e (k_{\rm i})+ N(p_N) \longrightarrow e'(k_{\rm f}) + K (q_K) + Y (p_Y)~.
\end{eqnarray}
To describe this process we make use of an isobar model, because
by utilizing this model the elementary amplitudes can be written in 
term of a frame independent operator which is 
required to include the Fermi motion in the nucleus. 
The process is schematically
shown in Fig.~\ref{fig:feynman}, where it is assumed that the 
electromagnetic interaction is mediated by one photon exchange.
The elementary transition operator can be written as
\begin{eqnarray}
  {M}_{\rm fi} &=& \epsilon_\mu\,J^\mu ~=~ 
  \bar{u}(\bvec{p}_Y)\,\sum_{i=1}^{6}\, 
  A_i(k^2,s,t,u)\,M_i\, u(\bvec{p}_N) ~.
\label{elementary_tr_op}
\end{eqnarray}
where  the virtual photon momentum 
$k=k_{\rm i}-k_{\rm f}$ and the Mandelstam variables are defined as
\begin{eqnarray}
s ~=~ (k+p_N)^2 ~~,~~ t ~=~ (k-q_K)^2 ~~,~~ u ~=~ (k-p_Y)^2 ~.
\label{mandelstam}
\end{eqnarray} 
The gauge and Lorentz invariant matrices $M_{i}$ in 
Eq.\,(\ref{elementary_tr_op}) are given by
\begin{eqnarray} \label{m1_el_op}
M_{1} &=& {\textstyle \frac{1}{2}}\, \gamma_{5}\, 
           \left( \epsilon \;\!\!\!\!/ \, k 
           \!\!\!/ - k \!\!\!/ \,\epsilon \;\!\!\!\!/ \right)~ ,\\
M_{2} &=& \gamma_{5}\, \left[\, (2q_K-k) \cdot \epsilon \; P \cdot k - 
          (2q_K-k) \cdot k \; P \cdot \epsilon \, \right]~ ,\\
M_{3} &=& \gamma_{5}\, \left(\, q_K\cdot k\; \epsilon \;\!\!\!\!/ - 
          q_K\cdot \epsilon\; k \!\!\!/ \,\right) ~ ,\\
M_{4} &=& i\, \epsilon_{\mu \nu \rho \sigma}\, \gamma^{\mu}\, q_K^{\nu}\,
          \epsilon^{\rho}\, k^{\sigma}~ ,\\
M_{5} &=& \gamma_{5} \left(\, q_K\cdot \epsilon \; k^{2} - q_K\cdot k\; 
          k \cdot \epsilon \,\right) ~ ,\\
M_{6} &=& \gamma_{5}\, \left(\, k \cdot \epsilon\; k \!\!\!/ - 
          k^{2} \epsilon \;\!\!\!\!/ \,\right)~ ,\label{m6_el_op}
\end{eqnarray}
with $P=\frac{1}{2}\, (\, p_N+p_Y\, )$ and $\epsilon_{\mu \nu \rho \sigma}$
represents the four dimensional Levi-Civita tensor with $\epsilon_{0123}=1$.
The coefficient functions $A_i$ are obtained from Feynman diagrams shown 
in Fig.~\ref{fig:feynman}.

For the purpose of the nuclear operator, the relativistic elementary operator
must be decomposed into its 'non-relativistic' form. In the case of free 
Dirac spinors, the operator in Eq.\,(\ref{elementary_tr_op}) can be 
decomposed into Pauli space 
\begin{eqnarray}
\lefteqn{ \overline{u}({\bvec p}_{Y})\,\sum_{i=1}^{6}\, 
A_{i}M_{i}\, u({\bvec p}_{N})
 ~=~ N\, \chi_{\rm f}^{\dagger}\, \biggl[\, {\cal{F}}_{1}
\,{\bvec{\sigma}} \cdot {\bvec{\epsilon}}
+ {\cal{F}}_{2}\,  {\bvec{\sigma}} \cdot {\bvec k}\, \epsilon_{0}
+ {\cal{F}}_{3}\, {\bvec{\sigma}} \cdot {\bvec k}\, 
{\bvec k} \cdot  {\bvec{\epsilon}} + {\cal{F}}_{4}\, 
{\bvec{\sigma}} \cdot {\bvec k}\, {\bvec p}_{N} \cdot 
{\bvec{\epsilon}} }\nonumber\\
 & & + {\cal{F}}_{5}\, {\bvec{\sigma}}
\cdot {\bvec k}\, {\bvec p}_{Y} \cdot {\bvec{\epsilon}}
+ {\cal{F}}_{6}\, {\bvec{\sigma}}\cdot {\bvec p}_{N}\,\epsilon_{0} 
+ {\cal{F}}_{7}\, {\bvec{\sigma}} 
\cdot {\bvec p}_{N}\, {\bvec k} \cdot {\bvec{\epsilon}}
+ {\cal{F}}_{8}\, {\bvec{\sigma}} \cdot {\bvec p}_{N}\, {\bvec p}_{N} 
\cdot {\bvec{\epsilon}} + {\cal{F}}_{9}\, 
{\bvec{\sigma}} \cdot {\bvec p}_{N}\, 
{\bvec p}_{Y} \cdot  {\bvec{\epsilon}}
\nonumber\\ 
&& + {\cal{F}}_{10}\, {\bvec{\sigma}} \cdot {\bvec p}_{Y}\, 
\epsilon_{0} 
+ {\cal{F}}_{11}\, {\bvec{\sigma}} \cdot {\bvec p}_{Y}\, {\bvec k} \cdot 
{\bvec{\epsilon}} 
+ {\cal{F}}_{12}\, {\bvec{\sigma}}
 \cdot {\bvec p}_{Y}\, {\bvec p}_{N} \cdot {\bvec{\epsilon}} + 
{\cal{F}}_{13}\, {\bvec{\sigma}} \cdot {\bvec p}_{Y}\, 
{\bvec p}_{Y} \cdot  {\bvec{\epsilon}}
\nonumber\\ 
 & &  + {\cal{F}}_{14}\, {\bvec{\sigma}} \cdot 
{\bvec{\epsilon}}\, {\bvec{\sigma}}
 \cdot {\bvec k}\, {\bvec{\sigma}} \cdot {\bvec p}_{N}
+ {\cal{F}}_{15}\, {\bvec{\sigma}} \cdot {\bvec p}_{Y}\, 
{\bvec{\sigma}} \cdot {\bvec{\epsilon}}\, {\bvec{\sigma}} \cdot {\bvec k}
 + {\cal{F}}_{16}\, {\bvec{\sigma}} \cdot {\bvec p}_{Y}\, 
{\bvec{\sigma}} \cdot {\bvec{\epsilon}}\, {\bvec{\sigma}} \cdot {\bvec p}_{N} 
\nonumber\\ 
 & & +\, {\cal{F}}_{17}\, {\bvec{\sigma}} \cdot {\bvec p}_{Y}\, 
 {\bvec{\sigma}} \cdot 
{\bvec k}\, {\bvec{\sigma}} \cdot {\bvec p}_{N}\, \epsilon_{0} 
+ {\cal{F}}_{18}\, {\bvec{\sigma}} \cdot {\bvec p}_{Y}\, 
{\bvec{\sigma}} \cdot {\bvec k}\, {\bvec{\sigma}} 
\cdot {\bvec p}_{N}\, {\bvec k} \cdot {\bvec{\epsilon}}
\nonumber\\ 
 & & + {\cal{F}}_{19}\, {\bvec{\sigma}} \cdot {\bvec p}_{Y}\, 
{\bvec{\sigma}} \cdot {\bvec k}\, {\bvec{\sigma}} 
\cdot {\bvec p}_{N}\, {\bvec p}_{N} \cdot  {\bvec{\epsilon}}
 +\, {\cal{F}}_{20}\, {\bvec{\sigma}} \cdot {\bvec p}_{Y}\, 
{\bvec{\sigma}} \cdot {\bvec k}\, {\bvec{\sigma}} 
\cdot {\bvec p}_{N}\, {\bvec p}_{Y} \cdot  {\bvec{\epsilon}}
\, \biggr] \, \chi_{\rm i} ~,
\label{nro1}
\end{eqnarray}
where 
\begin{eqnarray}
  N &=& \left(\frac{E_{N} + m_{N}}{2m_{N}} \right)^{\frac{1}{2}}
        \left(\frac{E_{Y} + m_{Y}}{2m_{Y}} 
        \right)^{\frac{1}{2}} ~, 
\end{eqnarray}
and the individual amplitudes ${\cal F}_{i}$ are given in 
Appendix~\ref{amplitudes_fi}

We will recast the elementary operator to a suitable form for
the nuclear process in the next section.
As shown in Ref.\,\cite{mart98} the terms of order $p^2/m^2$, i.e. 
$F_{16}$--$F_{20}$, can be dropped from the elementary operator,
since they come from the small spinor components. Furthermore, this will not
disturb the gauge invariance of the operator. Nevertheless, for the sake
of accuracy, the omission of these terms should be done carefully. Moreover,
unlike the situation in pion production, the particle momenta in our case
are always higher than those of pion.

\begin{figure}[!t]
  \begin{center}
    \leavevmode
    \epsfig{figure=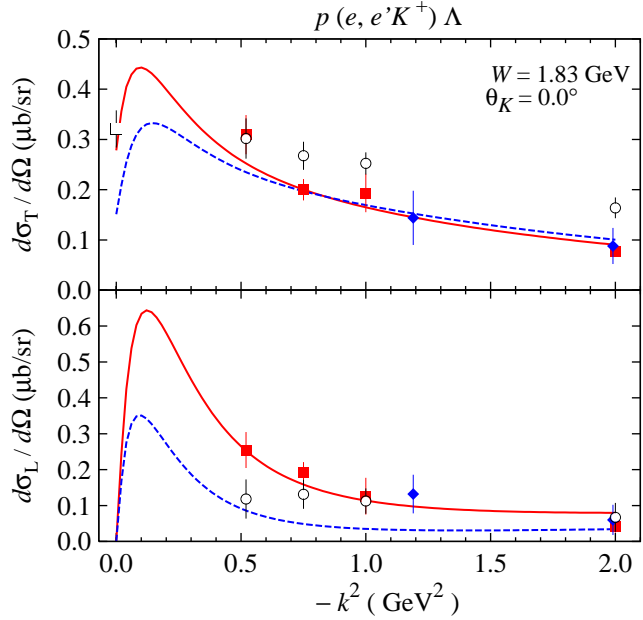,width=90mm}
    \caption{(Color online) Comparison between calculated cross
       sections obtained by including (solid lines) and excluding
       (dotted lines) the $D_{13}(1895)$ state with experimental
       data. Solid squares display the experimental measurement of 
       Niculescu {\it et al.}~\cite{niculescu}. This measurement
       has been reanalyzed by Mohring {\it et al.}~\cite{mohring}
       and shown here by the open circles. The solid
       diamonds are due to the old measurement
       by Brauel {\it et al.}~\cite{old_electro}.
       At photon point a photoproduction datum~\cite{photon_point}
       (open square) 
       is shown for comparison with the transverse cross section.}
   \label{fig:elementary_Q2} 
  \end{center}
\end{figure}

In this calculation we use the KAON-MAID parameterization~\cite{kaon-maid}.
The model consists of gauge-invariant background and resonances 
terms. The background terms include the standard $s$-,
$u$-, and $t$-channel contributions along with a contact term 
required to restore gauge invariance after hadronic form factors 
have been introduced~\cite{Haberzettl:1998eq}. 
The resonance part consists of three nucleon
resonances that have been found in the coupled-channels approach 
to decay into the $K\Lambda$ channel, i.e., the $S_{11}$(1650), 
$P_{11}$(1710), and $P_{13}(1720)$. Furthermore, the model also
includes  the $D_{13}(1895)$ state that is found to be important 
in the description of SAPHIR data~\cite{saphir}. 

\begin{figure}[!t]
  \begin{center}
    \leavevmode
    \epsfig{figure=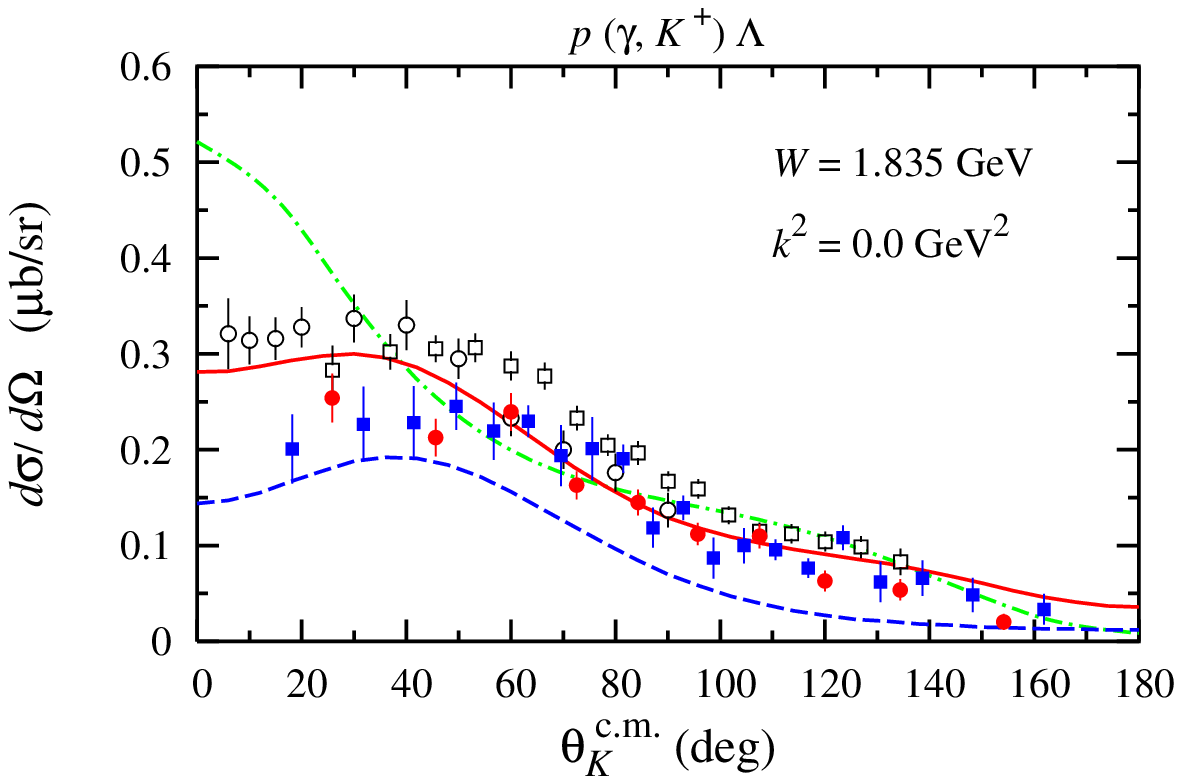,width=100mm}
    \caption{(Color online) As in Fig.~\protect\ref{fig:elementary_Q2},
    but for photoproduction ($k^2=0$). As a comparison, the calculated
    cross section from Ref.~\protect\cite{williams} is also shown by the
    dash-dotted line. Solid squares, solid circles represent the experimental
    data from Refs.~\cite{Glander:2003jw} and \cite{Tran:1998qw}, 
    respectively. Open squares and open circles represent 
    CLAS~\cite{Bradford:2005pt} and older
    data~\cite{old_data}, respectively.}
   \label{fig:elementary_photo} 
  \end{center}
\end{figure}

At finite $k^2$ the calculated transverse and longitudinal
cross sections obtained from this model are shown in 
Fig.~\ref{fig:elementary_Q2}. Since the model was fitted
to the data of Niculescu {\it et al.}~\cite{niculescu},
a sizeable discrepancy with the reanalyzed data~\cite{mohring}
appears in this figure. However, we note that the model
can also nicely describe the old measurement and photoproduction
data. As reported in Refs.~\cite{kaon-maid}, the inclusion of 
the $D_{13}(1895)$ state is important for the description of
the structure found in the $\gamma p\to K^+\Lambda$ total
cross section~\cite{saphir}. We will also investigate the 
influence of this state in the electroproduction of the 
hypertriton. To this end we show in Fig.~\ref{fig:elementary_Q2}
the calculated cross sections when this state were excluded.
Obviously, the magnitude of the cross sections is greatly
reduced once we omit this state, especially in the
case of the longitudinal one, where we can see from 
Fig.~\ref{fig:elementary_Q2} that at $k^2=-0.5$ GeV$^2$ the cross section 
is about four times smaller in this case.

In the case of photoproduction, sample of the angular distribution of
differential cross section is displayed in 
Fig.~\ref{fig:elementary_photo}, where we compare the prediction
of KAON-MAID and that obtained from Ref.~\cite{williams} with
experimental data from various measurements. It is obvious from this 
figure that there exist some discrepancies among the experimental data,
especially between the new SAPHIR~\cite{Glander:2003jw} 
and CLAS~\cite{Bradford:2005pt} data. The discrepancy and
its physics consequences have been thoroughly investigated in 
Ref.~\cite{Mart:2006dk} by means of a multipole model. 
In spite of this problem, however, Fig.~\ref{fig:elementary_photo} 
indicates that KAON-MAID still gives a reliable prediction for kaon
photoproduction. This becomes more obvious when we
compare its prediction with the prediction of 
Ref.~\cite{williams}, where the latter clearly
overestimates the experimental data at the very forward kaon angle.
Incidentally, in this region the result of the hypernuclear production is found
to be very sensitive to the elementary operator model used~\cite{Bydzovsky:2006wy}.

\section{The Nuclear Operator and Cross Sections}\label{sec:nuclear_op}
\begin{figure}[!h]
  \begin{center}
    \leavevmode
    \epsfig{figure=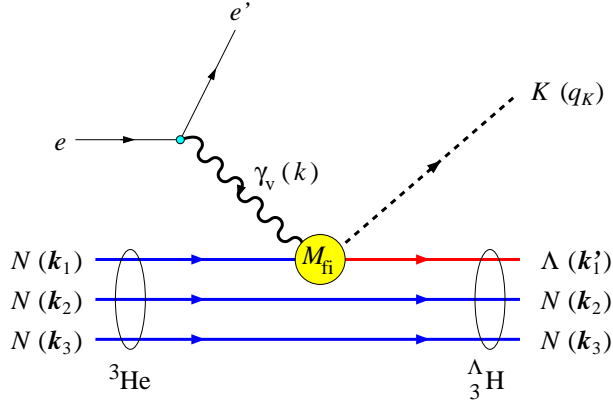,width=80mm}
    \caption{(Color online) Electroproduction of the hypertriton
        on a $^3$He target 
        in the impulse approximation, where the virtual photon
        interacts with only one nucleon inside the $^3$He. 
        The elementary operator
        $M_{\rm fi}=\epsilon_\mu\,J^\mu$ is given in Fig.~\protect\ref{fig:feynman}
        and Eq.~(\protect\ref{elementary_tr_op}). }
   \label{fig:impulse} 
  \end{center}
\end{figure}
In analogy to the case of photoproduction \cite{mart98}, 
we write the nuclear transition matrix
element in the laboratory frame as (see Fig.~\ref{fig:impulse})
\begin{eqnarray}
\langle\, {\rm f} \left| \, J^\mu \, \right| {\rm i} \,\rangle 
 &=& \sqrt{3}\,\int d^{3}\bvec{p}~d^{3}\bvec{q}~ \Psi_{\rm f}^{*}(\bvec{p}, 
 \bvec{q}') ~ J^\mu\, (k^0,\bvec{k},k_1^0,\bvec{k}_{1}, k_1'^0,\bvec{k}_1',
  q_K^0,\bvec{q}_K) ~
 \Psi_{\rm i}(\bvec{p},\bvec{q})~ ,~ 
\label{trans1}
\end{eqnarray}
where the integrations are taken over the three-body momentum coordinates
\begin{eqnarray}
  \bvec{p} ~=~ \frac{1}{2}\,(\bvec{k}_2 - \bvec{k}_3) ~~ , ~~
  \bvec{q} ~=~ \bvec{k}_1 ~ ,
\end{eqnarray}
and the hyperon momentum in the hypertriton is given by
\begin{eqnarray}
  \bvec{q}' &\equiv& \bvec{k}_1' ~=~\bvec{k}_1 + \frac{m_2+m_3}{m_1+m_2+m_3}\,\bvec{Q},
\end{eqnarray}
with the momentum transfer $\bvec{Q}=\bvec{k}-\bvec{q}_K$.
The factor of
$\sqrt{3}$ on the right hand side of Eq.\,(\ref{trans1}) comes from the 
anti-symmetry of the initial state. The derivation of this factor is given in
Appendix~\ref{derivation_sqrt3}. Note that in the following we will also
use the notations $\bvec{p}_N\equiv \bvec{k}_1$ and  $\bvec{p}_Y\equiv \bvec{k}_1'$ 
in order to facilitate the discussion of the elementary operator.

The $^3$He wave functions may be written as 
\begin{eqnarray}
\Psi_{\rm i}(\bvec{p},\bvec{q}) & = & \sum_{\alpha=(LSJljT)} 
\phi_{\alpha}(p,q)~ \left|\,\{(LS)J,(l {\textstyle \frac{1}{2}})
j\} {\textstyle \frac{1}{2}} M_{\rm i}\ ~\right\rangle ~ \left|\,
(T{\textstyle \frac{1}{2}}) {\textstyle \frac{1}{2}} M_{t}~\right\rangle
\nonumber\\
&=&\sum_{\alpha =(LSJljT)}
\sum_{\scriptsize \left. \begin{array}{c} \\[-6ex]
m_{L}m_{S}m_{l}\\[-2ex] m_{s}m_{J}m_{j} \end{array} \right.}\phi_{\alpha}
(p,q)~ (Lm_{L}Sm_{S}|J m_{J})~ 
(l m_{l}{\textstyle \frac{1}{2}} m_{s}|j m_{j}) \nonumber\\
&& \hspace{3cm} \times (J m_{J} j m_{j}
|{\textstyle \frac{1}{2}} M_{\rm i}) \, 
Y^{L}_{m_{L}} (\bvec{\hat{p}})Y^{l}_{m_{l}}(\bvec{\hat{q}})
\chi^{S}_{m_{S}}\chi_{m_{s}}^{\frac{1}{2}}~ \left|\,
(T{\textstyle \frac{1}{2}}){\textstyle \frac{1}{2}} M_{t}~\right\rangle ~ ,
\label{wfhe3}
\end{eqnarray}
where we have used the notation of Ref.\,\cite{deshalit63} for
the Clebsch-Gordan coefficients. The hypertriton wave 
functions can also be written in the form of Eq.\,(\ref{wfhe3}).

In Eq.\,(\ref{wfhe3}) we have introduced $\alpha = (LSJljT)$ 
to shorten the notation, where $L$, $S$, and $T$ are the total angular 
momentum, spin, and isospin of the pair (2,3), while for particle (1) the 
corresponding quantum numbers are labeled by $l$, $\frac{1}{2}$, 
and $\frac{1}{2}$, respectively. Their quantum numbers, along with the 
probabilities for the 34 partial waves, are listed in 
Table~\ref{probability_wf}, where we have used the Nijmegen93 
version of the $^3$He wave functions \cite{nijmegen93} and
the advanced model for the hypertriton wave functions given
in Ref.\,\cite{miyagawa93}.
Clearly, most contributions will come 
from the second partial wave ($\alpha =2$), 
which corresponds to the $S$-wave with isospin 0.

\begin{table}[htbp]
\renewcommand{\arraystretch}{0.7}
\begin{center}
\caption{Quantum numbers and probabilities (in \%) of the $^3$He and 
the hypertriton wave functions. }
\label{probability_wf}
\begin{ruledtabular}
\begin{tabular}{cccccccrr}
\\[-2ex]
$~\alpha~$ &$~L~$ & $~S~$ & $~J~$ & $~l~$ & $~2j~$ & $~2T~$ &
$P(^3{\rm He})$&$P(^3_{\Lambda}{\rm H})$ \protect\cite{miyagawa93}\\
[1ex]
\hline\\[-2ex]
 1 & 0 & 0 & 0 & 0 & 1 & 1 & 44.580  &   -~~~   \\
 2 & 0 & 1 & 1 & 0 & 1 & 0 & 44.899  &  93.491 \\
 3 & 2 & 1 & 1 & 0 & 1 & 0 &  2.848  &   5.794 \\
 4 & 0 & 1 & 1 & 2 & 3 & 0 &  0.960  &   0.034 \\
 5 & 2 & 1 & 1 & 2 & 3 & 0 &  0.189  &   0.027 \\
 6 & 1 & 0 & 1 & 1 & 1 & 0 &  0.089  &   0.004 \\
 7 & 1 & 0 & 1 & 1 & 3 & 0 &  0.198  &   0.008 \\
 8 & 1 & 1 & 0 & 1 & 1 & 1 &  1.107  &   -~~~   \\
 9 & 1 & 1 & 1 & 1 & 1 & 1 &  1.113  &   -~~~   \\
10 & 1 & 1 & 1 & 1 & 3 & 1 &  0.439  &   -~~~   \\
11 & 1 & 1 & 2 & 1 & 3 & 1 &  0.064  &   -~~~   \\
12 & 3 & 1 & 2 & 1 & 3 & 1 &  0.306  &   -~~~   \\
13 & 1 & 1 & 2 & 3 & 5 & 1 &  1.018  &   -~~~   \\
14 & 3 & 1 & 2 & 3 & 5 & 1 &  0.024  &   -~~~   \\
15 & 2 & 0 & 2 & 2 & 3 & 1 &  0.274  &   -~~~   \\
16 & 2 & 0 & 2 & 2 & 5 & 1 &  0.425  &   -~~~   \\
17 & 2 & 1 & 2 & 2 & 3 & 0 &  0.122  &   0.024 \\
18 & 2 & 1 & 2 & 2 & 5 & 0 &  0.095  &   0.018 \\
19 & 2 & 1 & 3 & 2 & 5 & 0 &  0.205  &   0.053 \\
20 & 4 & 1 & 3 & 2 & 5 & 0 &  0.053  &   0.006 \\
21 & 2 & 1 & 3 & 4 & 7 & 0 &  0.126  &   0.010 \\
22 & 4 & 1 & 3 & 4 & 7 & 0 &  0.038  &   0.007 \\
23 & 3 & 0 & 3 & 3 & 5 & 0 &  0.005  &   0.001 \\
24 & 3 & 0 & 3 & 3 & 7 & 0 &  0.008  &   0.001 \\
25 & 3 & 1 & 3 & 3 & 5 & 1 &  0.051  &   -~~~   \\
26 & 3 & 1 & 3 & 3 & 7 & 1 &  0.045  &   -~~~   \\
27 & 3 & 1 & 4 & 3 & 7 & 1 &  0.008  &   -~~~   \\
28 & 5 & 1 & 4 & 3 & 7 & 1 &  0.074  &   -~~~   \\
29 & 3 & 1 & 4 & 5 & 9 & 1 &  0.178  &   -~~~   \\
30 & 5 & 1 & 4 & 5 & 9 & 1 &  0.006  &   -~~~   \\
31 & 4 & 0 & 4 & 4 & 7 & 1 &  0.053  &   -~~~   \\
32 & 4 & 0 & 4 & 4 & 9 & 1 &  0.059  &   -~~~   \\
33 & 4 & 1 & 4 & 4 & 7 & 0 &  0.011  &   0.004 \\
34 & 4 & 1 & 4 & 4 & 9 & 0 &  0.009  &   0.003 \\[1ex]
\end{tabular}
\end{ruledtabular}
\end{center}
\end{table}

The elementary operator $J^\mu = (J^0, \bvec{J})$ is obtained from 
Eq.\,(\ref{nro1}), i.e.
\begin{eqnarray}
  J^0 &=& N~ \Bigl\{\, i{\cal{F}}_{17}\,\bvec{p}_{N} 
    \cdot (\bvec{p}_Y \times \bvec{k}) + ({\cal{F}}_{2}
    - \bvec{p}_{N} \cdot \bvec{p}_Y\,{\cal{F}}_{17}) \,
    \bvec{\sigma}\cdot\bvec{k} 
    + ({\cal{F}}_{6}+ \bvec{p}_{Y} \cdot \bvec{k}\, {\cal{F}}_{17})\, 
    \bvec{\sigma}\cdot\bvec{p}_{N} \nonumber\\
    && \hspace{7mm} 
    +\, ({\cal{F}}_{10}+\bvec{p}_{N} \cdot \bvec{k}\, {\cal{F}}_{17}) \,
    \bvec{\sigma}\cdot\bvec{p}_Y \Bigr\} ~,
    \label{el_J0}
\end{eqnarray}
and
\begin{eqnarray}
  \bvec{J} &=& -N~ \Bigl[~ ( {\cal{F}}_{1} + {\cal{F}}_{14}\; 
  \bvec{p}_{N} \cdot \bvec{k}-{\cal{F}}_{15}\;\bvec{p}_{Y} \cdot \bvec{k}
  -{\cal{F}}_{16}\; \bvec{p}_{N} \cdot \bvec{p}_{Y} )~
  \bvec{\sigma} \nonumber\\
  && \hspace{7mm}
  +~ \bvec{\sigma}\cdot\bvec{k}~~~
  \{~ ({\cal{F}}_{3}-
  \bvec{p}_{N}\cdot\bvec{p}_Y\; {\cal{F}}_{18})\; \bvec{k}
  + ( {\cal{F}}_{4}-{\cal{F}}_{14}-\bvec{p}_{N}\cdot\bvec{p}_Y\;
  {\cal{F}}_{19})\;\bvec{p}_{N} \nonumber\\
  && \hspace{22mm}
  +~ ( {\cal{F}}_{5}+{\cal{F}}_{15}-\bvec{p}_{N}\cdot\bvec{p}_Y\;
  {\cal{F}}_{20})\;\bvec{p}_{Y} ~\} \nonumber\\ 
  && \hspace{7mm}
  +~ \bvec{\sigma}\cdot\bvec{p}_N\,
  \{~ ({\cal{F}}_{7}+{\cal F}_{14} + 
  \bvec{p}_{Y}\cdot\bvec{k}\; {\cal{F}}_{18})\; \bvec{k}
  + ( {\cal{F}}_{8}+\bvec{p}_{Y}\cdot\bvec{k}\;
  {\cal{F}}_{19})\;\bvec{p}_{N} \nonumber\\
  && \hspace{22mm}
  +~ ( {\cal{F}}_{9}+{\cal{F}}_{16}+\bvec{p}_{Y}\cdot\bvec{k}\;
  {\cal{F}}_{20})\;\bvec{p}_{Y} ~\} \nonumber\\ 
  && \hspace{7mm}
  +~ \bvec{\sigma}\cdot\bvec{p}_Y\,
  \{~ ({\cal{F}}_{11}+{\cal F}_{15} + 
  \bvec{p}_{N}\cdot\bvec{k}\; {\cal{F}}_{18})\; \bvec{k}
  +~ ( {\cal{F}}_{12}+{\cal F}_{16}+\bvec{p}_{N}\cdot\bvec{k}\;
  {\cal{F}}_{19})\;\bvec{p}_{N} \nonumber\\
  && \hspace{22mm}
  +~ ( {\cal{F}}_{13}+\bvec{p}_{N}\cdot\bvec{k}\;
  {\cal{F}}_{20})\;\bvec{p}_{Y} ~\} \nonumber\\ 
  && \hspace{7mm}
  +~ i\; \{ -{\cal{F}}_{14}\; \bvec{p}_N\times\bvec{k} - {\cal{F}}_{15}\;
  \bvec{p}_Y\times\bvec{k} 
  + {\cal{F}}_{16}\, \bvec{p}_{N}\times\bvec{p}_Y \nonumber\\
  && \hspace{17mm}
  +~ \bvec{p}_N\cdot (\bvec{p}_Y \times
  \bvec{k})\; ({\cal{F}}_{18}\; \bvec{k} +
  {\cal{F}}_{19}\;\bvec{p}_{N} + {\cal{F}}_{20}
  \;\bvec{p}_Y\,)~\}~ \Bigr] ~.
  \label{el_Jmu}
\end{eqnarray}
It is obvious from Eqs.\,(\ref{m1_el_op})--(\ref{m6_el_op}) that the
gauge invariance of the elementary operator 
relates Eq.\,(\ref{el_J0}) and Eq.\,(\ref{el_Jmu}) by
\begin{eqnarray}
  J_0 &=& \bvec{k}\cdot \bvec{J} \,/\, k_0 ~,
  \label{current_conserv}
\end{eqnarray}
which slightly simplifies the numerical calculation since we can eliminate 
either $J^0$ or $J_z$ by  $\bvec{k}\cdot \bvec{J}=|\bvec{k}|J_z$.

For the purpose of calculating the observables it is
useful to rewrite the elementary operator in the form of a matrix
$[j]$, through the relation $J^\mu=[\sigma]\,[j]$, i.e.,
\begin{eqnarray}
  J^\mu &=& (1,\sigma_x,\sigma_y,\sigma_z) 
        \left(\begin{array}{cccc}
        j_{00} & j_{x0} & j_{y0} & j_{z0}\\
        j_{0x} & j_{xx} & j_{yx} & j_{zx}\\
        j_{0y} & j_{xy} & j_{yy} & j_{zy}\\
        j_{0z} & j_{xz} & j_{yz} & j_{zz}
        \end{array} \right)
        ~,
\label{eq:define_j}
\end{eqnarray}
where the individual components are given in 
Appendix~\ref{component_of_jmu}.

Since the hypertriton has isospin 0, we may drop the isospin part of  
the wave functions. 
By inserting the two nuclear wave functions in Eq.\,(\ref{trans1}) 
and writing symbolically ${\sf m} = (m_{L}m_{S}m_{l}m_{s}m_{J}m_{j})$ 
for the sake of brevity, we 
can recast the transition matrix element in the form of
\begin{eqnarray}
  \langle\, {\rm f} \left| \, J^\mu \, \right| {\rm i} \,\rangle &=& 
  \sqrt{3}\, \sum_{\alpha , \alpha '}
  \sum_{{\sf mm}'}
        \left(Lm_{L}Sm_{S}|Jm_{J}\right) \, 
        \left(Lm_{L}Sm_{S}|J'm_{J'}\right) \, 
        \left(lm_{l}{\textstyle \frac{1}{2}}m_{s}|jm_{j}\right)
   \nonumber\\
&& \times\,\left(l'm_{l'}{\textstyle \frac{1}{2}}m_{s'}|j'm_{j'}\right) \, 
        \left(Jm_{J}jm_{j}|{\textstyle \frac{1}{2}}M_{\rm i}\right) \, 
        \left(J'm_{J'}j'm_{j'}|{\textstyle \frac{1}{2}}M_{\rm f}\right) 
        \nonumber\\
&& \times\,\delta_{LL'}\,\delta_{m_{L}m_{L'}}\,
           \delta_{SS'}\,\delta_{m_{S}m_{S'}}\,\delta_{T0}
   \nonumber\\
&& \times\,\int p^{2}dp~d^{3}\bvec{q}\;\phi_{\alpha '}(p,q')\;
   \phi_{\alpha}(p,q)\; Y^{l'}_{m_{l'}}(\bvec{\hat q}')
   \, Y^{l}_{m_{l}}(\bvec{\hat q})\; \langle {\textstyle \frac{1}{2}},m_{s'}\, 
   |\, J^{\mu} \,|\, {\textstyle \frac{1}{2}},m_{s} \rangle \, ,
\label{trans2}
\end{eqnarray}
where we have performed the integration over the spectator solid angle,
\begin{eqnarray}
\int d\hat{\bvec{p}}~Y^{L'*}_{m_{L'}}(\hat{\bvec{p}})\,
Y^{L}_{m_{L}}(\hat{\bvec{p}}) 
&=& \delta_{LL'}\,\delta_{m_{L}m_{L'}}~,
\end{eqnarray}
as the relative momentum of the two spectators $\bvec{p}$ 
does not change. By using
\begin{eqnarray}
  \label{eq:jmu_def}
  J^\mu &=& j^\mu_0 + \sigma_x\, j^\mu_x + \sigma_y\, j^\mu_y +
            \sigma_z\, j^\mu_z \nonumber\\
        &=& \sum_{n=0,1}\;\sum_{m_n=-n}^{+n}\, 
            (-1)^{m_n}\,\sigma^{(n)}_{-m_n}\,
            [\,j^{\mu}\,]^{(n)}_{m_n} ~,
\end{eqnarray}
where the components of 
$[\,j^{\mu}\,]^{(n)}_{m_n}$ are given in 
Eq.~(\ref{eq:define_j}) and Appendix~\ref{component_of_jmu},
with
\begin{eqnarray}
  \left[\, j^{\mu}\, \right]^{(0)} &=& j^\mu_0~,\\
  \left[\, j^{\mu}\, \right]^{(1)}_{\pm 1} &=& 
  \mp\frac{1}{\sqrt{2}}(j^{\mu}_x\pm ij^{\mu}_y) ~,\\
  \left[\, j^{\mu}\, \right]^{(1)}_{0} &=& 
  j^{\mu}_z ~,\\
  \sigma^{(0)}&=&1~,\\
  \sigma^{(1)}&=&\bvec{\sigma}~,
\end{eqnarray}
and
\begin{eqnarray}
  \langle {\textstyle \frac{1}{2}},m_{s'}\,|\,\sigma_{-m_n}^{(n)}\,|\, 
  {\textstyle \frac{1}{2}},m_{s}\rangle &=& 
  \sqrt{2}\, (-1)^{n-\frac{1}{2}-m_{s'}+m_{n}}
    \left( {\textstyle \frac{1}{2}}-\! m_{s'}
           {\textstyle \frac{1}{2}}m_{s}|n m_{n} \right) ~,
\label{bracket1}
\end{eqnarray}
we can rewrite
Eq.\,(\ref{trans2}) in the form of
\begin{eqnarray}
  \langle\, {\rm f} \left| \, J^\mu \, \right| {\rm i} \,\rangle &=& 
  \sqrt{6}\, \sum_{\alpha , \alpha '}~
  \sum_{{\sf m},{\sf m}'}~  \sum_{n,m_{n}}
        \left(Lm_{L}Sm_{S}|Jm_{J}\right) \, 
        \left(Lm_{L}Sm_{S}|J'm_{J'}\right) \, 
        \left(lm_{l}{\textstyle \frac{1}{2}}m_{s}|jm_{j}\right)
   \nonumber\\
&& \times\,\left(l'm_{l'}{\textstyle \frac{1}{2}}m_{s'}|j'm_{j'}\right) \, 
        \left(Jm_{J}jm_{j}|{\textstyle \frac{1}{2}}M_{\rm i}\right) \, 
        \left(J'm_{J'}j'm_{j'}|{\textstyle \frac{1}{2}}M_{\rm f}\right) \,
        \left({\textstyle \frac{1}{2}}-\! m_{s'}
              {\textstyle \frac{1}{2}}m_{s} | n m_{n} \right)
        \nonumber\\
&& \times\, (-1)^{n-\frac{1}{2}-m_{s'}}
            \delta_{LL'}\,\delta_{m_{L}m_{L'}}\,
            \delta_{SS'}\,\delta_{m_{S}m_{S'}}\,\delta_{T0}
   \nonumber\\
&& \times\,\int p^{2}dp~d^{3}\bvec{q}\;\phi_{\alpha '}(p,q')\;
   \phi_{\alpha}(p,q)\; Y^{l'}_{m_{l'}}(\bvec{\hat q}')
   \, Y^{l}_{m_{l}}(\bvec{\hat q})
   \; \left[\,j^{\mu}\,\right]^{(n)}_{m_n}
    \, .
\label{trans3}
\end{eqnarray}
Note that the elementary operator $\left[\,j^{\mu}\,\right]^{(n)}_{m_n}$
is completely frame independent, since it is independent from the frame
where $\epsilon^\mu$ and $\sigma^{(n)}$ are defined.
Hence, by summing and averaging over the nuclear spins 
we can construct the spin averaged Lorentz tensor
\cite{tiator81} 
\begin{eqnarray}
  W^{\mu\nu} &=& \frac{1}{2} \sum_{M_{\rm i}M_{\rm f}}
  \langle\, {\rm f} \left| \, J^\mu \, \right| {\rm i} \,\rangle 
  \langle\, {\rm f} \left| \, J^\nu \, \right| {\rm i} \,\rangle ^* ~,
\end{eqnarray}
which is related to the nuclear structure functions by 
\begin{eqnarray}
  W_{\rm T} &=& \frac{1}{4\pi}~ (W_{xx}+W_{yy}) ~,\\
  W_{\rm L} &=& \frac{1}{4\pi}~ W_{00} ~,\\
  W_{\rm TT} &=& \frac{1}{4\pi}~ (W_{xx}-W_{yy}) ~,\\
  W_{\rm LT} &=& \frac{1}{4\pi}~ (W_{0x}+W_{x0}) ~.
\end{eqnarray}
The exclusive cross section $^3{\rm He}(e,e'K^+)^3_\Lambda{\rm H}$ can be
written as
\begin{eqnarray}
  \frac{d^5\sigma}{d\varepsilon_{\rm f}\, d\Omega_{e'}\, d\Omega_K} &=& 
  \Gamma~ \frac{d\sigma_v}{d\Omega_K} ~,
\end{eqnarray}
where the flux of virtual photons is given by
\begin{eqnarray}
 \Gamma &=&\frac{\alpha}{2\pi^2}~ 
 \frac{\varepsilon_{\rm f}}{\varepsilon_{\rm i}}
  ~ K_L ~ \frac{1}{-k^2}~ \frac{1}{1-\epsilon} ~,
\label{flux}
\end{eqnarray}
and the differential cross section for kaons produced by virtual photons
can be written as 
\begin{eqnarray}
  \label{eq:ds_domega_virtual}
  \frac{d\sigma_v}{d\Omega_K} &=& \frac{d\sigma_{\rm T}}{d\Omega_K} +
  \epsilon_{\rm L}~ \frac{d\sigma_{\rm L}}{d\Omega_K} +
  \epsilon~ \frac{d\sigma_{\rm TT}}{d\Omega_K}~ \cos 2\phi_K +
  \sqrt{2\epsilon_{\rm L}(1+\epsilon)}~ 
  \frac{d\sigma_{\rm LT}}{d\Omega_K}~ \cos \phi_K ~,
\end{eqnarray}
with the virtual photon polarization of
\begin{eqnarray}
  \epsilon &=& \left( 1-2 \,\frac{\bvec{k}^2}{k^2}\, \tan^2
                  {\textstyle \frac{1}{2}} \theta_e \right)^{-1} ~,
\end{eqnarray}
and 
\begin{eqnarray}
  \epsilon_{\rm L} &=& -\frac{k^2}{\bvec{k}^2}\,\epsilon  ~.
\end{eqnarray}
The cross sections are conventionally measured in the c.m.
system. In this frame of reference the individual cross sections are
given by
\begin{eqnarray}
  \frac{d\sigma_{\rm T}}{d\Omega_K^{\rm c.m.}} &=& \alpha_e \,
  \frac{q_{K}^{\rm c.m.}}{K_L}\, \frac{M_{^3_\Lambda\!{\rm H}}}{2W}\, W_{\rm T}^{\rm c.m.} ~ ,\\
  \frac{d\sigma_{\rm L}}{d\Omega_K^{\rm c.m.}} &=& \alpha_e \,
  \frac{q_{K}^{\rm c.m.}}{K_L}\, \frac{M_{^3_\Lambda\!{\rm H}}}{W}\,W_{\rm L}^{\rm c.m.} ~ ,\\
  \frac{d\sigma_{\rm TT}}{d\Omega_K^{\rm c.m.}} &=& \alpha_e \,
  \frac{q_{K}^{\rm c.m.}}{K_L}\, \frac{M_{^3_\Lambda\!{\rm H}}}{2W}\, W_{\rm TT}^{\rm c.m.} ~ ,\\
  \frac{d\sigma_{\rm LT}}{d\Omega_K^{\rm c.m.}} &=& -\alpha_e \,
  \frac{q_{K}^{\rm c.m.}}{K_L}\, \frac{M_{^3_\Lambda\!{\rm H}}}{2W}\, W_{\rm LT}^{\rm c.m.} ~ ,
\end{eqnarray}
where $\alpha_e=e^2/4\pi$ the fine structure constants and 
we have defined the photon equivalent energy [also in Eq.\,(\ref{flux})]
\begin{eqnarray}
  K_L &=& \frac{W^2-M_{\rm He}^2}{2M_{\rm He}} ~.
\end{eqnarray}
The transformation from the laboratory to c.m. frames affects the
longitudinal structure functions only and leaves the transverse ones
unchanged, i.e.,
\begin{eqnarray}
  W_{\rm T}^{\rm c.m.}  &=&  W_{\rm T}^{\rm lab}~, \\
  W_{\rm L}^{\rm c.m.}  &=&  W_{\rm L}^{\rm lab}\,{\bvec{k}_{\rm c.m.}^2}/{\bvec{k}^2}~, \\
  W_{\rm TT}^{\rm c.m.} &=&  W_{\rm TT}^{\rm lab}~,\\
  W_{\rm LT}^{\rm c.m.} &=&  W_{\rm LT}^{\rm lab}\,\sqrt{{\bvec{k}_{\rm c.m.}^2}/{\bvec{k}^2}}~.
\end{eqnarray}

\section{Results and Discussion}\label{sec:result}
The summations over ${\sf m}$ and ${\sf m}'$ in Eq.~(\ref{trans3}) are
significantly reduced by the properties of the Clebsch-Gordan
coefficient. As the result, we only need to sum over the angular-momentum 
and spin projections $m_{J},m_{J'},m_S$, and $m_s$, since 
the other projections are fixed by the relations
\begin{eqnarray}
  m_{s'}&=& m_s-m_n~,\\
  m_{j'}&=& M_{\rm f}-m_{J'}~,\\
  m_{j }&=& M_{\rm i}-m_{J}~,\\
  m_{L }&=& m_{J'}-m_S~,\\
  m_{l'}&=& m_{j'}-m_{s'}~,\\
  m_{l }&=& m_{j}-m_{s}~.
\end{eqnarray}

As the first step, we need to check our Fortran code. This has been
performed by calculating the elementary cross sections and comparing
the results with those obtained from the original elementary code. 
For this purpose we replace the wave functions in Eq.~(\ref{trans3}) 
by unity. As a consequence, Eq.~(\ref{trans3}) is greatly reduced to
\begin{eqnarray}
  \langle\, {\rm f} \left| \, J^\mu \, \right| {\rm i} \,\rangle &=& 
  \sqrt{2}\, \sum_{n,m_{n}} (-1)^{n-1/2-M_{\rm f}}
        \left({\textstyle \frac{1}{2}} \;-\!\!M_{\rm f}\; 
        {\textstyle \frac{1}{2}}\; M_{\rm i}|nm_n\right)
   \,\left[\,j^{\mu}\,\right]^{(n)}_{m_n}
    \, ,
\label{trans_elementary}
\end{eqnarray}
and the  transverse and longitudinal cross sections can expressed in terms of 
\begin{eqnarray}
\label{cs_elementary1}
  \frac{d\sigma_{\rm T}}{d\Omega_K^{\rm c.m.}} &=& 
  \frac{q_{K}^{\rm c.m.}}{k^{\rm c.m.}}\, \frac{m_pm_\Lambda}{32\pi^2W^2}\, 
        \sum_{i=0}^3 \left( |j_{ix}|^2+|j_{iy}|^2\right) ~ ,\\
  \frac{d\sigma_{\rm L}}{d\Omega_K^{\rm c.m.}} &=& 
  \frac{q_{K}^{\rm c.m.}}{k^{\rm c.m.}}\, \frac{m_pm_\Lambda}{32\pi^2W^2}\, 
        \sum_{i=0}^3  2|j_{i0}|^2 ~ ,
\label{cs_elementary2}
\end{eqnarray}
which can be shown to be identical with the standard definitions of
the transverse and longitudinal cross sections in the elementary process.
However, we do not use Eqs.~(\ref{cs_elementary1}) and (\ref{cs_elementary2})
to check the code. Instead, we calculate the elementary cross sections
by using the main code, that is used to compute the nuclear cross sections, 
but we replace the wave functions in Eq.~(\ref{trans3}) by unity.
The output of the Fortran code shows a precise agreement with the
cross sections calculated directly by using the CGLN amplitudes
\cite{chew} (i.e. the solid lines in Figs.~\ref{fig:elementary_Q2}
and \ref{fig:elementary_photo}), 
which is the standard way of calculating the cross sections in 
KAON-MAID. This result convinces us that our code has calculated
the cross sections properly.

\subsection{Photoproduction of the Hypertriton}
\label{sub_result_photo}
\begin{figure}[!t]
  \begin{center}
    \leavevmode
    \epsfig{figure=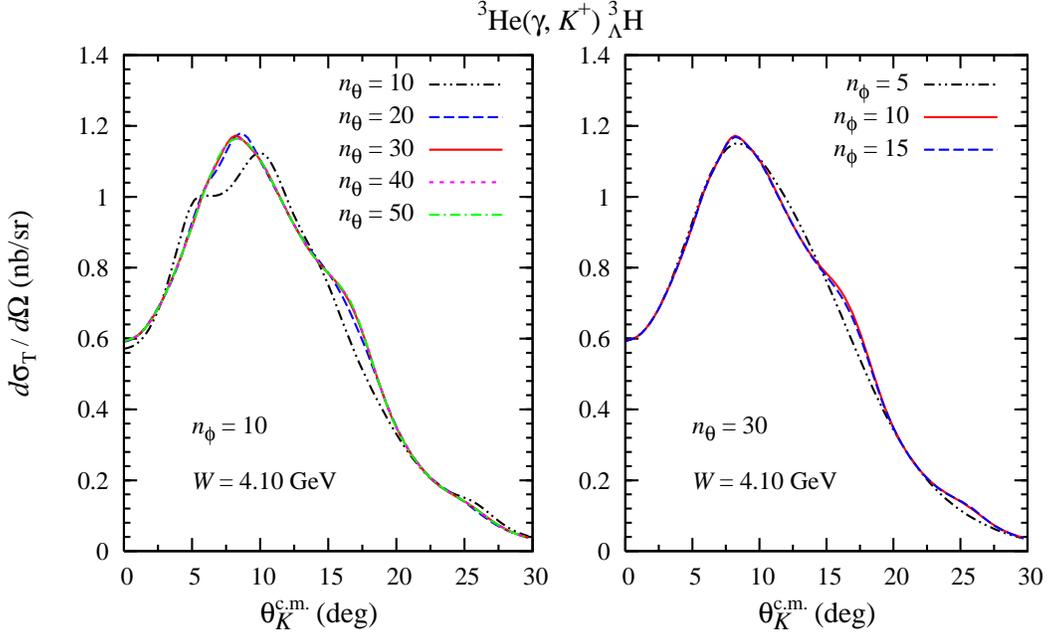,width=140mm}
    \caption{(Color online) Variations of the hypertriton photoproduction 
             cross section for different values of the number of 
             Gauss supporting points
             $n_\theta$ and $n_\phi$ in the angular integration of
             $d{\bvec{\hat q}}$. Only $s$-wave parts of the nuclear
             wave functions have been used to obtain these curves.}
   \label{fig:numerical_stability} 
  \end{center}
\end{figure}

As shown in Table~\ref{probability_wf} the initial $^3$He wave function
contains 34 components and the final hypertriton wave function contains
16 components. Since we use the impulse approximation, the quantum numbers
of the pair remain unchanged, which is represented by the three Kronecker
delta functions $\delta_{LL'}\delta_{SS'}\delta_{T0}$ in Eq.~(\ref{trans3}). 
This selection rule significantly 
reduces the number of non-zero diagonal and interference terms for the
components of wave functions from $34\times 16$ to just 64 components. 
In both wave functions the number of supporting points for the $p$
and $q$ momenta are 34 and 20, respectively.

To calculate the four-dimensional integrals in Eq.~(\ref{trans3}) we
have used Gaussian integration. We first carried out the overlap 
integral in $p$, because it is easier, and stored the result 
in a $20\times 20\times 64$ array, where the last component is
intended for the index of the non-zero overlap integrals.
Since the computation of the integrals 
is very time consuming, the number of the supporting points should be
limited as small as possible, without sacrifying the numerical stability
of the integration. To this end in Fig.~\ref{fig:numerical_stability} 
we display the variations of the cross sections as functions of the 
number of Gauss supporting points for the angular integration in 
Eq.~(\ref{trans3}), i.e., $n_\theta$ and $n_\phi$. It is obvious from this figure that the result
of integrations starts to become stable for $n_\theta\ge 30$ and $n_\phi\ge 10$.
Therefore, in the following discussion we shall only use the results with
$n_\theta=30$ and $n_\phi=10$. For the full calculation at every 
point of cross sections of interest we have carried out an integration over
$34\times 20\times 64\times 30\times 10=13056000$ grid points.
The numerical computation becomes more challenging because the integrand consists 
of the elementary operator $\left[\,j^{\mu}\,\right]^{(n)}_{m_n}$ in the form 
of $4\times 4$ complex-component 
matrix. Fortunately, current conservation given by Eq.~(\ref{current_conserv})
reduces the required information to $4\times 3=12$ components. The result,
which is equivalent to an integration over 156 millions grid points, 
is then summed over angular momentum and spin projections $m_{J},m_{J'},m_S$, and $m_s$,
as described above.

\begin{figure}[!t]
  \begin{center}
    \leavevmode
    \epsfig{figure=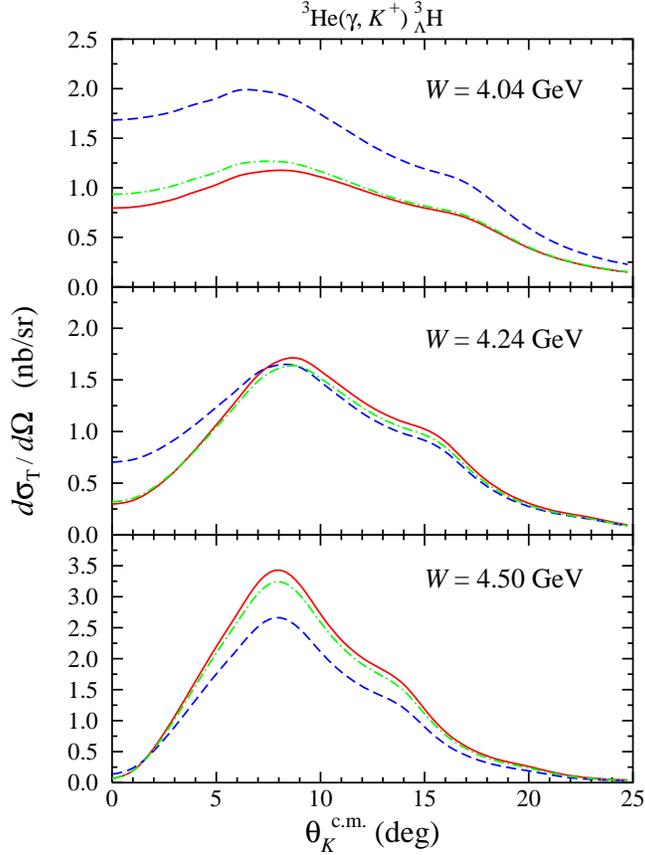,width=90mm}
    \caption{(Color online) Effects of Fermi motion on the differential 
      cross sections at three different total c.m. energies. The dashed curve 
      is obtained from the 
      ``frozen nucleon'' approximation ($\langle {\bvec k}_1 \rangle = 0$),
      the dash-dotted curve is obtained with an average momentum of 
      $\langle {\bvec k}_1 \rangle = -\frac{1}{3}{\bvec Q}$, while the solid
      curve shows the exact treatment of Fermi motion.}
   \label{fig:frozen} 
  \end{center}
\end{figure}

As in the previous study \cite{mart98} we have also investigated 
contribution of non-localities generated by Fermi motion in the 
initial and final nuclei. The exact treatment of Fermi motion is included 
in the integrations over the  wave functions in Eq.~(\ref{trans3}), whereas
a local approximation can be carried out by freezing the operator 
at an average nucleon momentum 
\begin{eqnarray}
\label{local}
\langle {\bvec k}_1 \rangle &=& -\kappa \frac{A-1}{2A}{\bvec Q} 
~=~ -\frac{\kappa}{3}\bvec{Q}~, 
\end{eqnarray} 
since $A=3$. For $\kappa =0$, Eq.~(\ref{local}) 
corresponds to the ``frozen nucleon'' approximation, whereas  
$\kappa =1$  yields the average momentum approximation. 

Figure \ref{fig:frozen} displays the effect of Fermi motion on the differential 
cross sections at three different total c.m. energies. Note that these energies
correspond to the photon lab energies $E_\gamma=1.5$ GeV, 1.8 GeV, and 2.2 GeV,
used in our previous work \cite{mart98} for making the comparison easier. 
Reference~\cite{tiator2} has shown that the effect of Fermi motion in the pion
photoproduction  in the $s$- and $p$-shells is in part 
simulated by the average momentum assumption 
$\langle {\bvec k}_1 \rangle = -\frac{1}{3}{\bvec Q}$. Figure~\ref{fig:frozen} 
obviously shows that this phenomenon is also found in the hypertriton photoproduction,
whereas the use of ``frozen nucleon'' approximation ($\langle {\bvec k}_1 \rangle = 0$)
leads to significantly different results. Although 
the average momentum assumption can approximate the exact treatment of
Fermi motion, for the sake of accuracy 
we will use the exact treatment of Fermi motion in the following discussion.

\begin{figure}[!t]
  \begin{center}
    \leavevmode
    \epsfig{figure=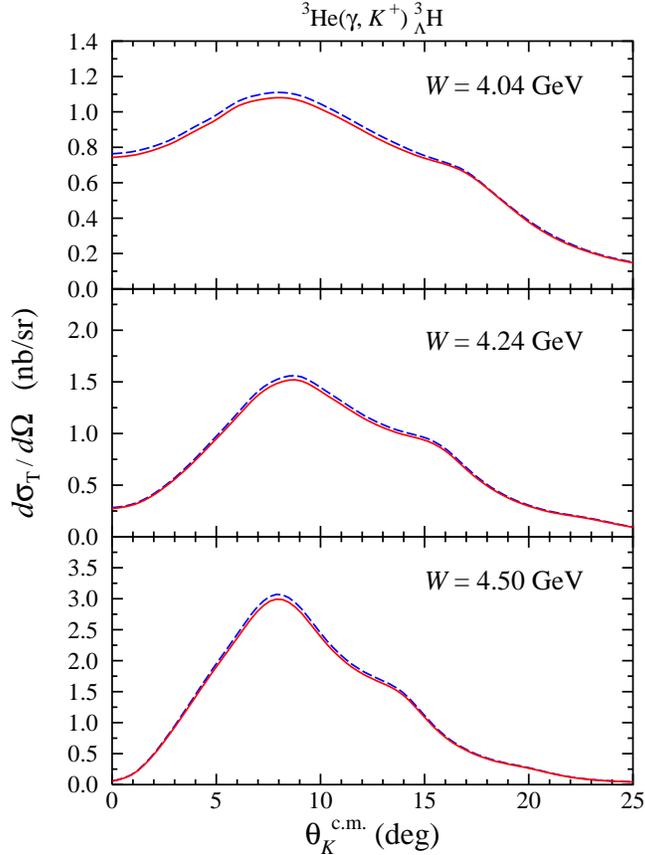,width=90mm}
    \caption{(Color online) The effect of the Coulomb correction on the differential
    cross sections at three different total c.m. energies. The dashed curves show
    the results without Coulomb correction, whereas the solid lines are obtained after
    including this correction.}
   \label{fig:coulomb} 
  \end{center}
\end{figure}

In the final state the positively charged kaon interacts with
the hypertriton by means of the Coulomb force. Therefore, in
our calculation a Coulomb correction factor must be taken into 
account. For this purpose we follow Ref.~\cite{tia_th}, who 
introduced the Gamow factor 
\begin{eqnarray}
  \label{coulomb}
  F_G (|\bvec{q}_K|) &=& \frac{2\pi\zeta}{{\rm exp}[2\pi\zeta] - 1} ~,
\end{eqnarray}
with 
\begin{eqnarray}
  \zeta &=& \frac{\alpha_e}{v_K} ~=~ \alpha_e~\frac{E_K}{|\bvec{q}_K|} ~,
\end{eqnarray}
and $\alpha_e = e^2/4\pi$ to account for the Coulomb effect in pion 
photoproduction off $^3$He at threshold. In Ref.~\cite{tia_th} 
it has been shown that this factor is important to help to describe 
experimental data at threshold. 
In the case of hypertriton production this factor is found to be
negligible, as shown in Fig.~\ref{fig:coulomb}. The same finding
has been also reported by the previous study~\cite{mart98}. This
result can be understood, because the corresponding photon energy 
in the present work (as well as in Ref~\cite{mart98}) is much higher
than the threshold energy of pion photoproduction on $^3$He.

\begin{figure}[!t]
  \begin{center}
    \leavevmode
    \epsfig{figure=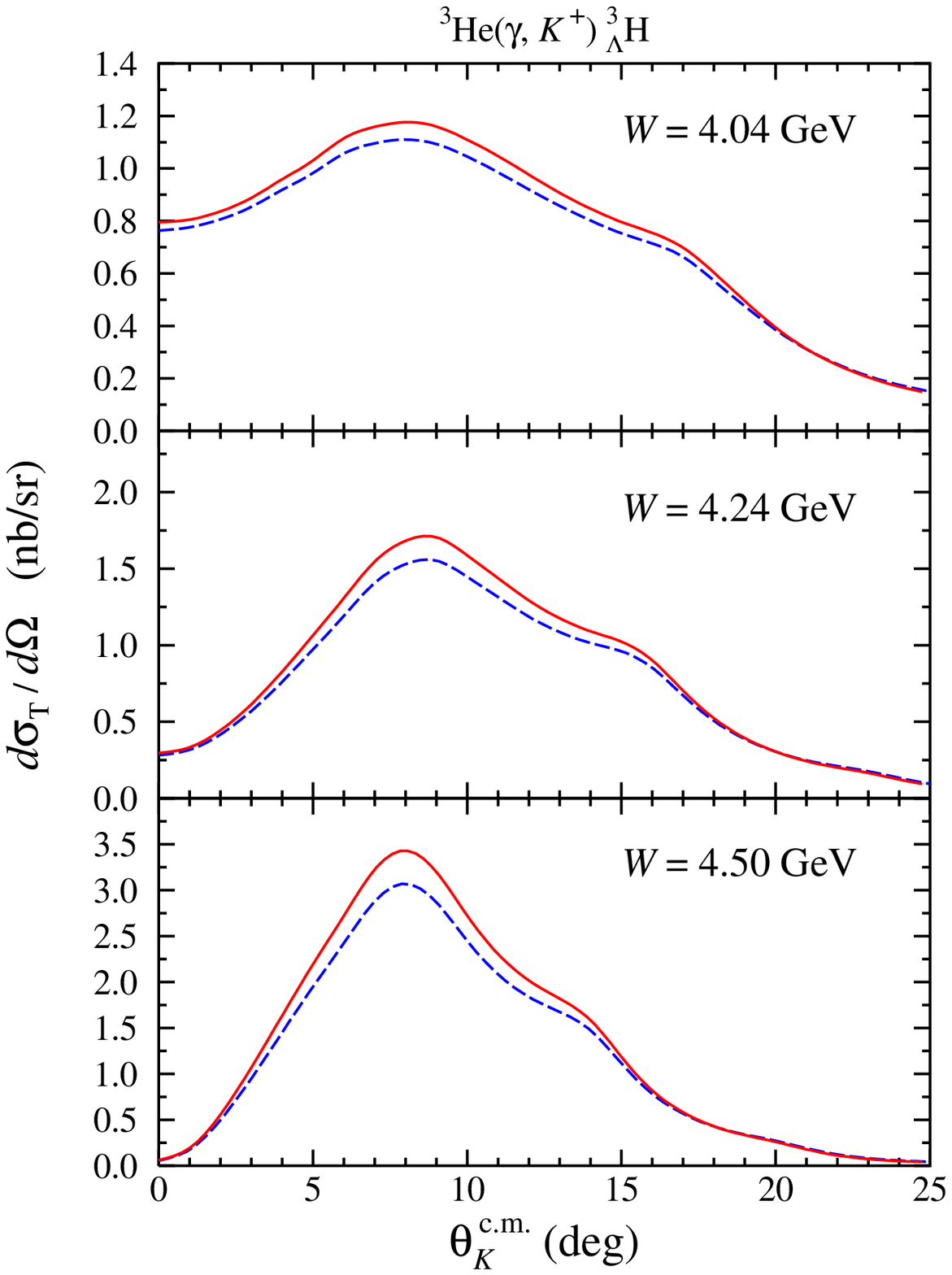,width=90mm}
    \caption{(Color online) Effects of the higher partial waves in the 
      $^3{\rm He}(\gamma,K^+)^3_\Lambda{\rm H}$ process at three different total c.m.
      energies. Dashed curves are the cross sections obtained by using 
      only $s$-wave, while solid curves
      exhibit the results after including all partial waves listed in 
      Table~\protect\ref{probability_wf}. }
   \label{fig:varwave} 
  \end{center}
\end{figure}

The influence of higher partial waves on the cross section is shown in 
Fig.~\ref{fig:varwave}. As shown in this figure the effect is only essential
at the cross section bumps ($\theta_K^{\rm c.m.}\approx 8^\circ$), while at
very small (and very large) kaon scattering angle the effect vanishes. 
Nevertheless, for the sake of accuracy, the following results have been 
obtained from calculations by using all available partial waves 
given in Table~\ref{probability_wf}. 

\begin{figure}[!t]
  \begin{center}
    \leavevmode
    \epsfig{figure=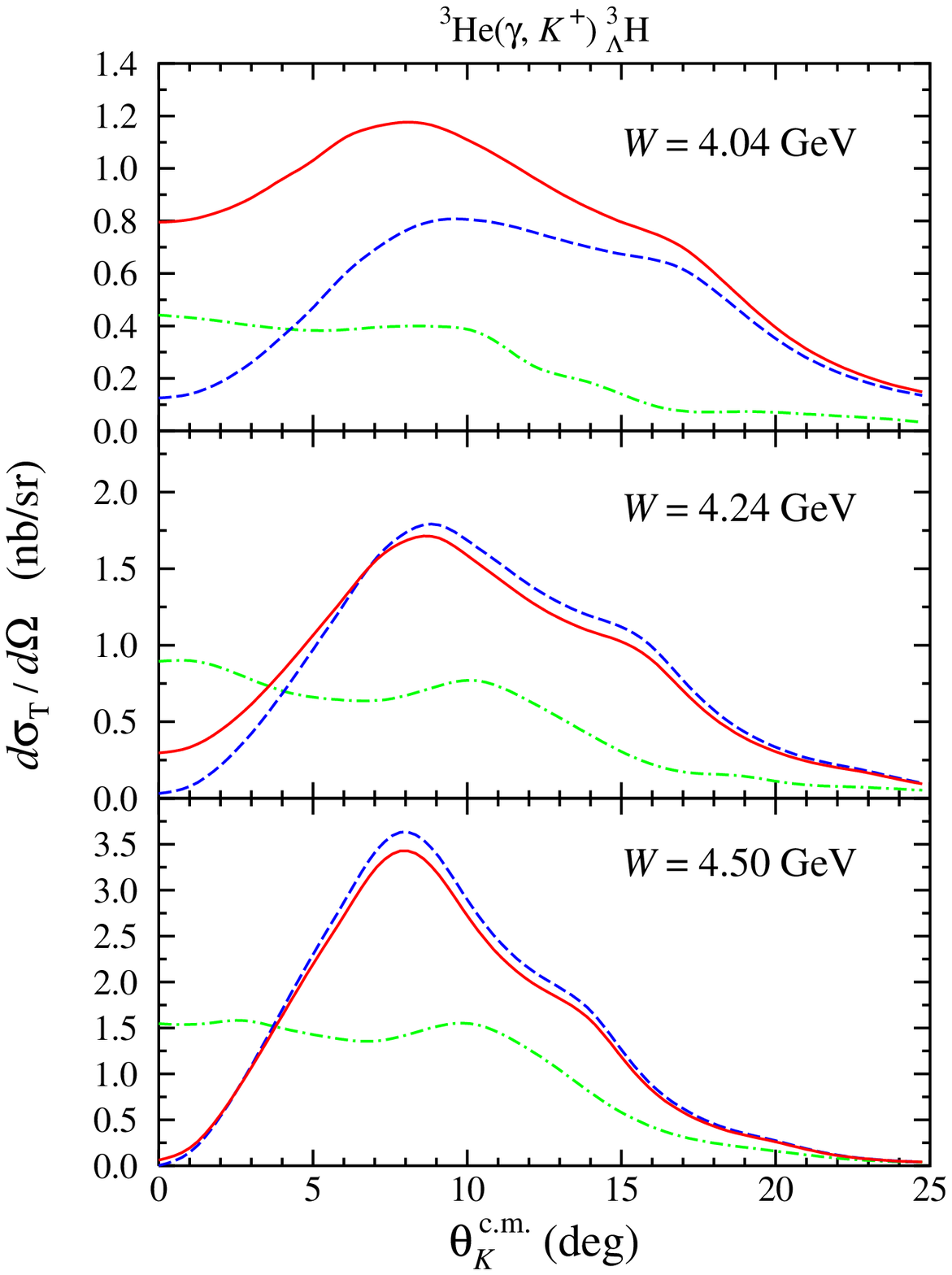,width=90mm}
    \caption{(Color online) Comparison between differential cross sections of 
      hypertriton photoproduction obtained in the present work with the missing
      resonance $D_{13}(1895)$ included in the elementary operator (solid lines) and
      those obtained by excluding this resonance (dashed lines). Results from 
      the previous work \protect\cite{mart98} which have been obtained by using different
      nuclear wave functions and different elementary operator~\cite{williams} are 
      shown by the dash-dotted lines. In all cases all available partial waves
      have been taken into account.}
   \label{fig:varmod} 
  \end{center}
\end{figure}

As in the elementary process, 
contribution of the missing resonance $D_{13}(1895)$ in the hypertriton production
is also found to be significant, especially at $W=4.04$ GeV (see Fig.~\ref{fig:varmod}). 
We feel that this is reasonable, because 
the energy corresponds to the elementary total c.m. energy 
$W_{\gamma p\to K\Lambda}=1.9$ GeV, i.e., almost at the resonance
pole position. As shown in Fig.~\ref{fig:varmod}, the effect gradually disappears
at higher energies. We note that, due to the strong nuclear suppression at large kaon scattering
angles, this effect also vanishes for $\theta_K^{\rm c.m.} > 25^\circ$.
Therefore, $W\approx 4.04$ and 
$0^\circ \lesssim\theta_K^{\rm c.m.} \lesssim 20^\circ$ 
represent an example of the recommended kinematics for the measurement of
hypertriton photoproduction. This conclusion is apparently also supported 
by Fig.~\ref{fig:frozen}, where for this kinematics the variation of 
differential cross sections due to the effect of non-localities 
is found to be remarkable.

From  Fig.~\ref{fig:varmod}
we also observe that the present calculation yields considerable discrepancy 
with the result of the previous calculation~\cite{mart98}. We estimate that
this discrepancy originates from the different nuclear wave functions and
elementary operator used. Previous calculation~\cite{mart98} used the 
wave function of $^3$He obtained as a 
solution of the Faddeev equations with the Reid soft core potential 
\cite{kim}, and the simple hypertriton wave function developed in Ref. 
\cite{congleton} that consists of only two partial waves. 
Furthermore, we note that in Ref.~\cite{mart98}
the calculated differential cross section would increase by a factor of
about three if only $s$-wave were used (see Fig. 5 
of Ref.~\cite{mart98}), in spite of the fact that 
contribution from other partial waves is less than 6\% (see Table I
of Ref.~\cite{mart98}). In conclusion we would like to say that
present calculation provides a more reliable result, since it uses
more accurate nuclear wave functions~\cite{nijmegen93,miyagawa93} 
and elementary operator~\cite{kaon-maid},
while the effect of the high-momentum partial waves
seems to be more reasonable.

\begin{figure}[!t]
  \begin{center}
    \leavevmode
    \epsfig{figure=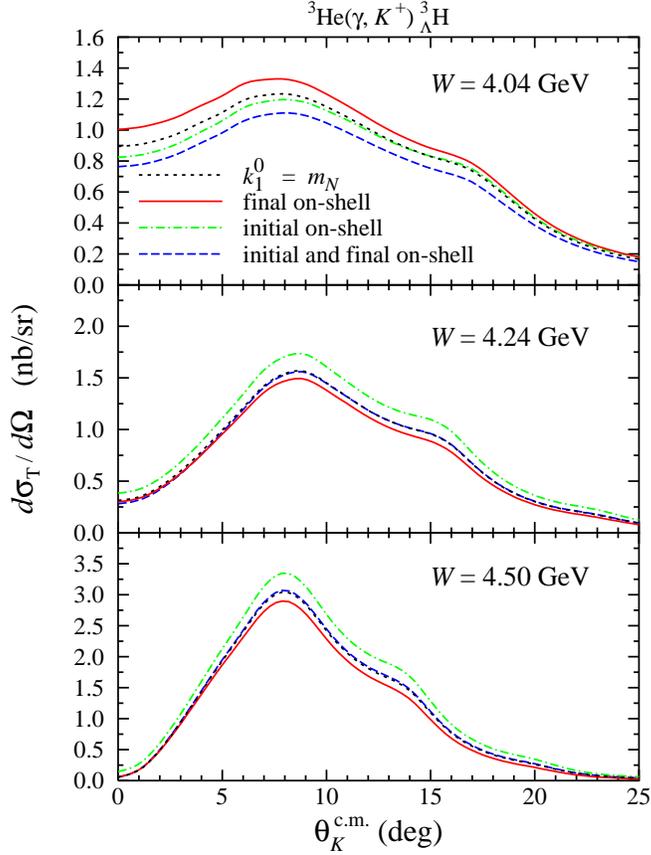,width=90mm}
    \caption{(Color online) The effect of different off-shell assumptions on the 
      differential cross section calculated at three different total c.m. energies.
      The dash-dotted curves illustrate the calculation with the initial nucleon in
      $^3$He on-shell, i.e., $\left[ k_1^0 = (m_N^2+{\bvec k}_1^2)^{1/2} \right]$, 
      while solid curves are obtained by assuming the final hyperon in 
      the hypertriton to be on-shell
      $\left[ k_{1}'^0 = (m_Y^2+{\bvec k}_{1}'^2)^{1/2} \right]$.
      The dashed curves show the calculation with both initial and final baryons 
      are on-shell, while the dotted curves are obtained by assuming $k_1^0 = m_N$
      (both initial and final baryons are off-shell).}
   \label{fig:offshell} 
  \end{center}
\end{figure}

It is obvious that all baryons involved in the hypertriton productions are 
off-shell. The elementary operator has been, however, constructed and fitted
to experimental data where both initial nucleon and final hyperon are on-shell.
In view of this, it is of interest to study the influence of the off-shell 
behavior of these baryons on the calculated cross sections. For this purpose
we make four assumptions:
\begin{enumerate}
\item Both initial and final baryons are on-shell,
  \begin{eqnarray}
    \label{eq:assum1}
    k_1^0 = (m_N^2+{\bvec k}_1^2)^{1/2}~~~,~~~
    k_{1}'^0 = (m_Y^2+{\bvec k}_{1}'^2)^{1/2}~.
  \end{eqnarray}
\item The initial nucleon is on-shell and the final hyperon is off-shell, 
  \begin{eqnarray}
    \label{eq:assum2}
  k_1^0 = (m_N^2+{\bvec k}_1^2)^{1/2} ~~~,~~~
  k_{1}'^0 = k_1^0+k_0-E_K~.
\end{eqnarray}
\item The initial nucleon is off-shell and the final hyperon is on-shell,
  \begin{eqnarray}
    \label{eq:assum3}
  k_{1}^0 = k_1'^0+E_K-k_0 ~~~,~~~  k_{1}'^0 = (m_Y^2+{\bvec k}_{1}'^2)^{1/2}~.
\end{eqnarray}
\item Both initial and final baryons are off-shell. In this case the static approximation, 
  \begin{eqnarray}
    \label{eq:assum4}
  k_1^0=m_N ~~~,~~~ k_{1}'^0 = k_1^0+k_0-E_K~,
\end{eqnarray}
is used.
\end{enumerate}
Note that we have used the first assumption in the previous figures for the sake
of simplicity. 
These four different off-shell assumptions result in complicated variations of the
differential cross sections as depicted in Fig.~\ref{fig:offshell}. At $W=4.04$ GeV
the first assumption yields the smallest cross section, while the third assumption
leads to the largest cross section. However, at higher $W$ the situation changes,
the latter gives in fact the smallest cross sections.  At $W=4.04$ GeV we
estimate that experimental data 
at forward angles with about 10\% error bars would be able to check these off-shell 
assumptions. For other kinematics (higher $W$) the cross section differences are
presumably to small in view of the present technology~\cite{Dohrmann:2004xy}. 

\subsection{Electroproduction of the Hypertriton}
It has been widely known that electroproduction process offers more possibilities
to study the structure of nucleons and nuclei. Furthermore, Ref.~\cite{tiator81}, 
e.g., has shown that electroproduction of pion off $^3$He reveals different phenomena, 
compared to photoproduction of pion on $^3$He. The effects of Fermi motion, for instance,
is found to be more profound in electroproduction rather than in photoproduction. 
The effects of different off-shell assumptions are also found to be more considerable
in electroproduction, especially in the transverse cross section. Motivated by
these observations, here we continue our investigation described in 
Subsection~\ref{sub_result_photo} to the finite $k^2$ regions.

\begin{figure}[!t]
  \begin{center}
    \leavevmode
    \epsfig{figure=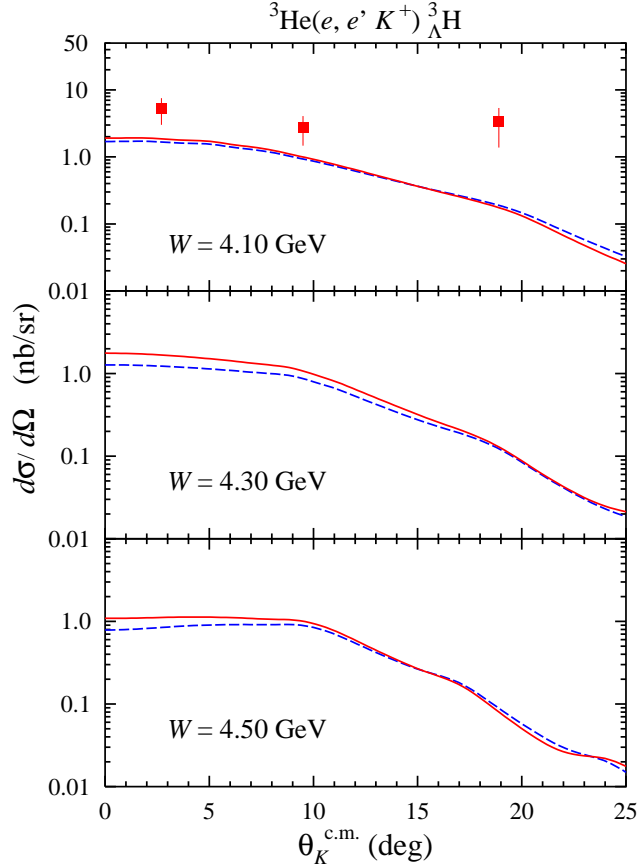,width=90mm}
    \caption{(Color online) The effect of the $s$-wave approximation on the
      differential cross section of the hypertriton electroproduction. The dashed curves
      are the cross sections obtained by using only $s$-wave, while the solid curves
      exhibit the results after including all partial waves listed in 
      Table~\protect\ref{probability_wf}. The curves have been obtained from
      Eq.~(\ref{eq:ds_domega_virtual}) with 
      $k^2=-0.35$ GeV$^2$ and $\epsilon=0.762$, whereas the azimuth angle $\phi$ 
      has been averaged. Experimental data are from
      Ref.~\protect\cite{Dohrmann:2004xy}.}
   \label{fig:k2vsth_full} 
  \end{center}
\end{figure}

\begin{figure}[!t]
  \begin{center}
    \leavevmode
    \epsfig{figure=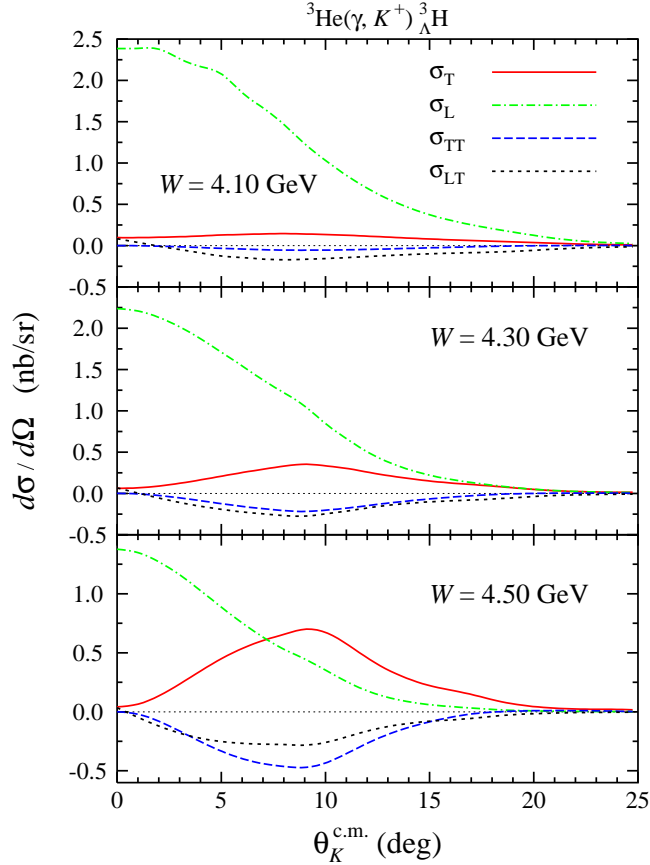,width=90mm}
    \caption{(Color online) Individual differential cross sections, $d\sigma_{\rm T}/d\Omega$,
     $d\sigma_{\rm L}/d\Omega$, $d\sigma_{\rm TT}/d\Omega$, and $d\sigma_{\rm LT}/d\Omega$,
     as a function of the kaon scattering angle at three different total c.m. energies.}
   \label{fig:cross_all} 
  \end{center}
\end{figure}

The result for hypertriton electroproduction shows, however, different behavior
compared to the case of pion electroproduction of $^3$He. This is demonstrated 
by Fig.~\ref{fig:k2vsth_full}, where we can see that the effect of the 
higher partial waves is considerably smaller than in the case
of photoproduction (cf. Fig.~\ref{fig:varwave}). Only at higher $W$
and very forward directions, where the momentum transfer $\bvec{Q}$ is
significantly large, the effects are sizable.
Note that the electroproduction
cross sections exhibit quite different shapes compared to the photoproduction
ones. This indicates that the longitudinal terms dominate other
contributions in all three kinematics shown in  Fig.~\ref{fig:k2vsth_full}.
This conjecture is proven by Fig.~\ref{fig:cross_all}, from which it is obvious that
the fall-off structure of the longitudinal cross sections drives the whole
shapes of the cross section shown in Fig.~\ref{fig:k2vsth_full}, whereas
the behavior of the transverse cross sections (with peaks at 
$\theta_K^{\rm c.m.}\approx 8^\circ$) is similar to that of the photoproduction
cross sections given in Fig.~\ref{fig:varwave}. 

We have found that the dominant behavior of the longitudinal
cross sections originate from the missing resonance $D_{13}(1895)$.
Excluding this resonance results in a reduction of the longitudinal
cross sections by one order of magnitude, whereas the transverse
ones decreases only by a factor of two. This result indicates that
the behavior of the longitudinal terms in the elementary operator
(see Fig.~\ref{fig:elementary_Q2}) is persistent and even gets amplified
in the nuclear cross sections. 

\begin{figure}[!t]
  \begin{center}
    \leavevmode
    \epsfig{figure=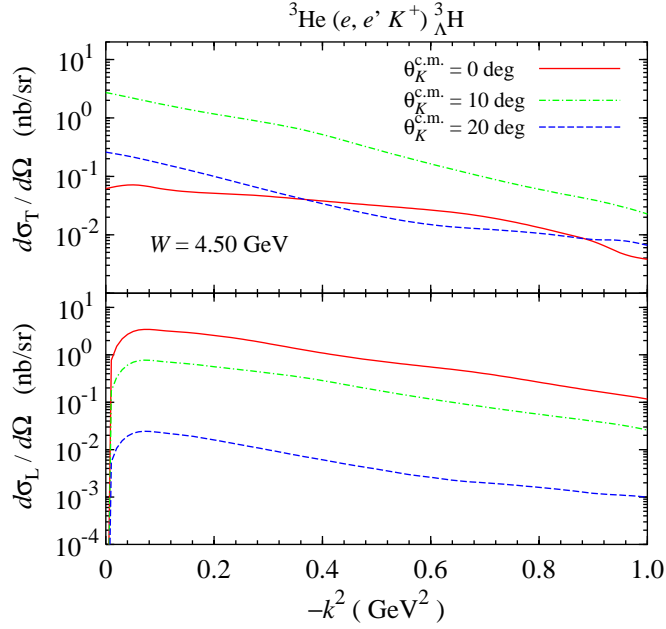,width=90mm}
    \caption{(Color online) Transverse and longitudinal cross sections as a function
    of the virtual photon momentum squared at three different kaon scattering angles.}
   \label{fig:k2_profile} 
  \end{center}
\end{figure}

The $k^2$ evolutions of both transverse and longitudinal cross sections
are displayed in Fig.~\ref{fig:k2_profile}. Both  Figs.~\ref{fig:cross_all} and
\ref{fig:k2_profile} reveals the different behaviors of the longitudinal
and transverse (along with other) cross sections at the forward direction 
and at $\theta_K^{\rm c.m.}\approx 10^\circ$. The result demonstrates the 
possibility to isolate the longitudinal cross section from other
contributions by measuring the process at forward directions. On the other hand, 
measurements at $\theta_K^{\rm c.m.}\approx 10^\circ$ (with averaged $\phi$) 
can give us the transverse cross section.

\begin{figure}[!t]
  \begin{center}
    \leavevmode
    \epsfig{figure=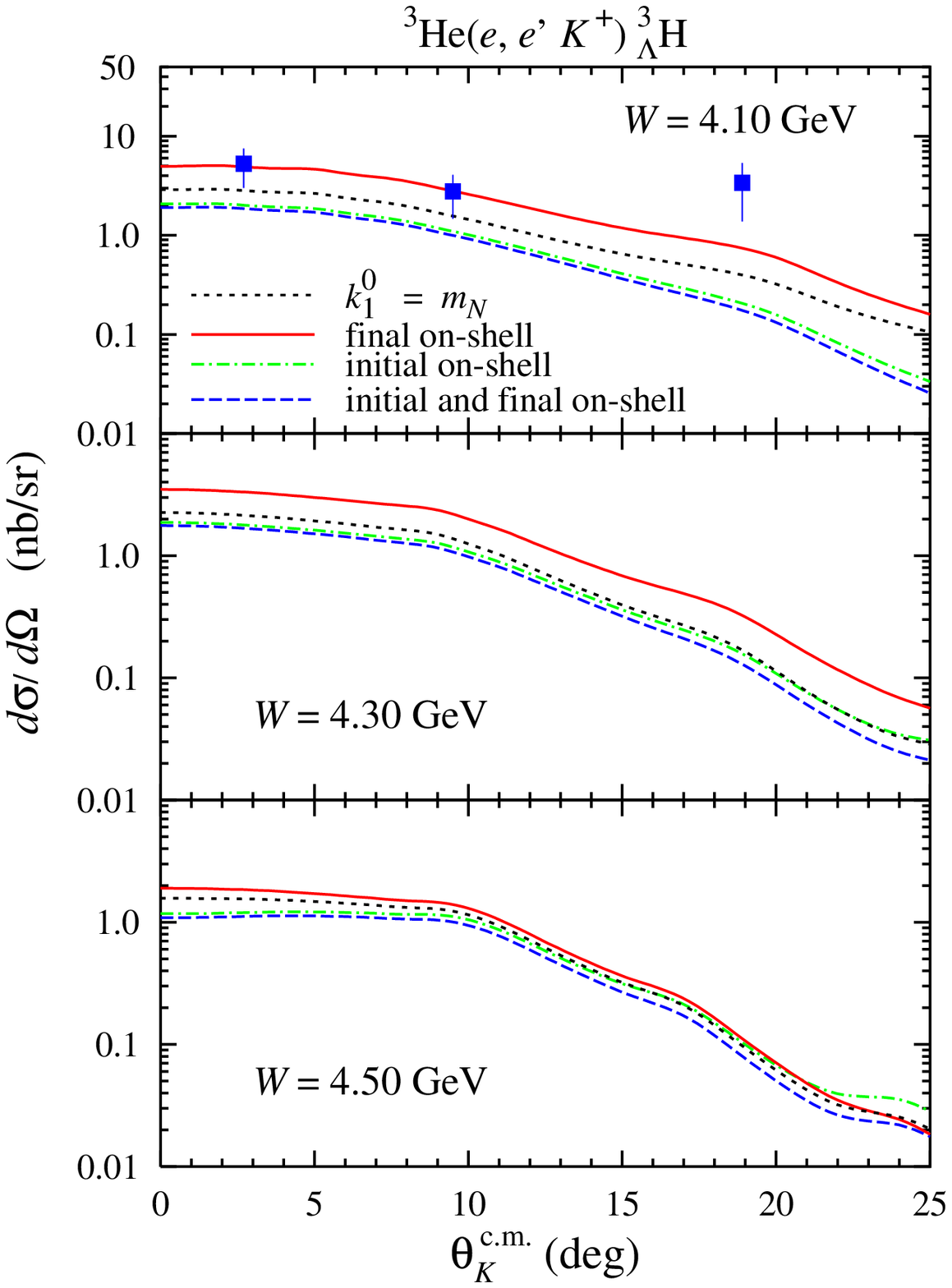,width=90mm}
    \caption{(Color online) Off-shell effect in the hyperon electroproduction.
    Notations for the curves and data are as in Fig.~\protect\ref{fig:offshell}.}
   \label{fig:k2_off} 
  \end{center}
\end{figure}

Off-shell effects are also found to be important in the case of
electroproduction. This fact is clearly exhibited in Fig.~\ref{fig:k2_off}, 
where we can see that the ``final hyperon on-shell'' assumption can
nicely shift the cross section upward to reproduce the experimental data
at $\theta_K^{\rm c.m.}=2.7^\circ$ and $9.5^\circ$. Obviously, this
finding is in contrast to the phenomenon observed in the pion
photoproduction off $^3$He~\cite{tiator2}, where the assumption
that the initial nucleon is on-shell yields a better agreement
with experimental data. However, the present finding can be understood 
as follows: The hyperon binding energy in the hypertriton 
is much weaker than the binding energy of the nucleon in the $^3$He.
Therefore, shifting the hyperon in the final state closer to its 
mass-shell moves the model closer to reality. The experimental data 
point at $\theta_K^{\rm c.m.}=18.9^\circ$ seems, however, 
to be very difficult to reproduce. Although the elementary cross 
section at this kinematics slightly increases, the nuclear suppression 
from the two nuclear wave functions is sufficiently strong to reduce 
the cross sections at  $\theta_K^{\rm c.m.}\ge 10^\circ$.

\begin{figure}[!t]
  \begin{center}
    \leavevmode
    \epsfig{figure=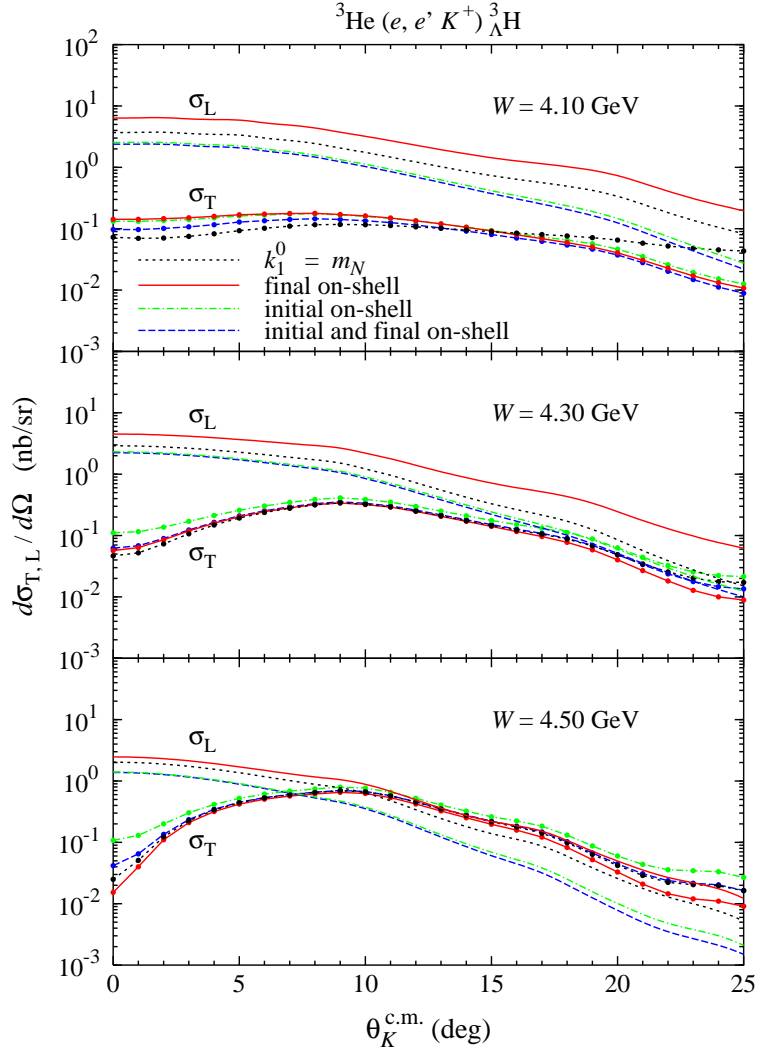,width=100mm}
    \caption{(Color online) Comparison between off-shell effects on
        the longitudinal and on the transverse cross sections at
        different values of $W$ and $k^2=-0.35$ GeV$^2$. Note that
        for the sake of visibility the transverse cross sections are
        plotted with lines and points, while the convention for the
        line types is as in the longitudinal cross sections.}
   \label{fig:offshell_el} 
  \end{center}
\end{figure}

At $W=4.30$ GeV and  $W=4.50$ GeV the various off-shell assumptions yield 
quite different
phenomena compared to those in the case of photoproduction (compare the two
lower panels of Fig.~\ref{fig:k2_off} and Fig.~\ref{fig:offshell}). In the
case of photoproduction the assumption that the final hyperon is on-shell
yields the smallest cross sections, whereas the situation is opposite in
the case of electroproduction. Again, this behavior originates from the
longitudinal terms. As shown by Fig.~\ref{fig:offshell_el}, 
for all $W$ shown, the longitudinal cross sections are larger than the
transverse ones. In the former, the off-shell effects are more 
profound and the assumption that the hyperon in the final state is
on-shell always yields the largest cross section. Such behavior does not
show up in the transverse cross sections. In fact, by comparing 
the transverse cross sections shown in Fig.~\ref{fig:offshell_el} 
and in Fig.~\ref{fig:offshell} we can clearly see that the result
presented here is still consistent with that of the photoproduction.

We have also found that although the phenomenon of the 
dominant longitudinal cross sections 
originates from the contribution of the missing resonance
$D_{13}(1895)$, the fact that the final-hyperon-on-shell 
assumption always yields the largest cross section is not
affected by the omission of this resonance. 

\subsection{Future Consideration}

\begin{figure}[!t]
  \begin{center}
    \leavevmode
    \epsfig{figure=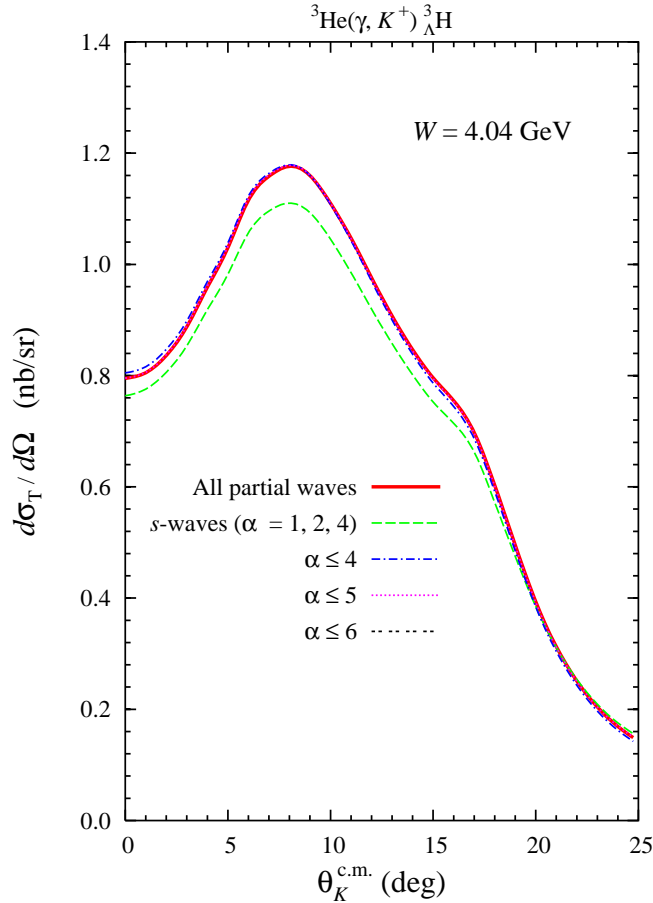,width=90mm}
    \caption{(Color online) Variations of the hypertriton photoproduction 
             cross section at $W=4.04$ GeV as functions of
             the number of partial waves used and the kaon c.m. angle.
             Note that the cross sections for $\alpha\le 5$ and $\alpha\le 6$
             coincide with that of the full calculation (using all partial waves).
            }
   \label{fig:var_high_wave} 
  \end{center}
\end{figure}

The massive numerical integrations in the full calculation 
using all partial waves described above requires special
attention in the future. One way to reduce this task is
by limiting the number of the involved elementary amplitudes 
${\cal F}_i$ given in Appendix~\ref{amplitudes_fi}. 
It has been shown in Ref.~\cite{Adelseck:1986fb} that to a good
approximation the ``big-big parts'' of the Dirac
spinors (in our case, the ${\cal F}_{16}$ -- ${\cal F}_{20}$)
of a special isobar model can be safely neglected. 
However, the discrepancies between the full calculation and
this approximation depends critically on both photon and
nucleon energies (see, for instance, Fig.~2 of Ref.~\cite{Adelseck:1986fb}).
Thus, careful inspections in a wide range
of kinematics should be previously performed, before we can
apply this approximation in the hypertriton production.

Another method which might be of interest is by limiting the number of the partial
waves used in the calculation. 
As shown in Fig.~\ref{fig:varwave}, the discrepancy
between the results of the full calculation and the $s$-wave approximation
can reach about 10\%. Therefore, the use of only $s$-wave
would not be recommended for a precise calculation. However, 
the ultimate question is: What is the minimum value of $\alpha$, for which
we would obtain the best approximation? To answer this question 
we have calculated the cross sections at $W=0.4$ GeV
by using different numbers of partial waves, from the $s$-wave
approximation up to $\alpha=6$,
and we demonstrate the result in Fig.~\ref{fig:var_high_wave}. From
this figure we can immediately conclude that the calculation with
$\alpha\le 4$ would provide a good approximation, whereas
by using $\alpha\le 5$ we could achieve the best result.
We note that in the latter the number of non-zero diagonal 
and interference terms of the components of the wave functions
turns out to be 8. Obviously, this method provides a significant
CPU-time reduction compared to the full calculation which
has 64 non-zero components. In spite of this encouraging
result, however, an extensive investigation of this approximation
in a wide range of kinematics will need to be addressed 
in the future, before we can draw a  firm conclusion that
we really need the triton and hypertriton wave functions with only
five partial waves to obtain a precise calculation
of the hypertriton production.

\section{Conclusions and Outlook}\label{sec:conclusion}
We have investigated the photo- and electroproduction
of hypertriton on the $^3$He nucleus by utilizing the
modern nuclear wave functions obtained as a solution 
of the Faddeev equations and the elementary operator
KAON-MAID. It has been shown that the proper treatment 
of the Fermi motion is essential in this process. While
the average momentum approximation  
$\langle {\bvec k}_1 \rangle = -\frac{1}{3}{\bvec Q}$
can partly simulate the Fermi motion, the ``frozen nucleon''
assumption yields very different results, especially at
lower energies. This indicate that the effect of
non-localities generated by Fermi motion is important
in the electromagnetic production of hypertriton.
On the other hand, although the exited meson is 
a positive kaon, the Coulomb effect induced by
its interaction with the hypertriton 
is found to be negligible. The influence of higher
partial waves is also found to be small, in contrast
to the finding in the previous work. The off-shell
assumption is found to be more important in the
case of electroproduction rather than in photoproduction.
Our finding indicates that the available experimental
data favor the assumption that the initial nucleon
is off-shell, whereas the final hyperon is on-shell.
This seems to be reasonable, since the hyperon
in the hypertriton is less bound than the nucleon
in the initial $^3$He nucleus. The longitudinal cross
sections are dominant in the electroproduction process.
This originates from the longitudinal terms of the
missing resonance $D_{13}(1895)$ in the elementary
operator. Nevertheless, the influence of various
off-shell assumptions on the longitudinal cross 
sections is not affected by the exclusion
of this resonance. Experimental measurements are strongly
required, especially in the case of photoproduction,
where we can partly settle the problems of the elementary
operator due to the lack of data consistency and of the
knowledge on the nucleon resonances as well as 
hadronic coupling constants. For this case, $W\approx 4.04$ 
and forward directions represent the recommended
kinematics, for which the effects of non-localities, 
missing resonance $D_{13}(1895)$, as well as 
various off-shell assumptions are found to be quite
significant.
Further measurements of the hypertriton electroproduction are
obviously useful, especially if we want to explore the
role of the longitudinal terms in the elementary
operator and to recheck the trend of the angular distribution
of the differential cross sections.

\section*{Acknowledgment}
The authors thank K. Miyagawa for providing the hypertriton
and $^3$He wave functions and W. Gl\"ockle for explaining
the normalization of the hypertriton wave function.
T.M. thanks the Physics Department of the 
Stellenbosch University for the hospitality
extended to him during his stay in Stellenbosch, where
part of this work was carried out. T.M. also 
acknowledges the support from the University of Indonesia.
The work of B.I.S.v.d.V has been supported by the South 
African National Research Foundation under Grant number 
GUN 2048567.


\appendix

\section{The Elementary Amplitudes ${\cal F}_i$ }
\label{amplitudes_fi}
Here we list the elementary amplitudes  ${\cal F}_1-{\cal F}_{20}$
defined by Eq.~(\ref{nro1}). Note that all energies and three-momenta are 
given in the $\gamma-N$ c.m. or lab system.
\begin{eqnarray}
{\cal{F}}_{1} & = & k_{0}A_{1} + k \cdot q_K A_{3} + \{2P \cdot k - 
k_{0}(m_{N} + m_{Y})\}A_{4} - k^{2}A_{6} ~,\\
{\cal{F}}_{2} & = & -A_{1} - E_K A_{3} - (E_{N} + E_{Y} - m_{N} - 
m_{Y})A_{4} + k_{0}A_{6} ~, \\
{\cal{F}}_{3} & = & A_{3} - A_{6} ~, \\
{\cal{F}}_{4} & = & A_{3} + A_{4} ~, \\
{\cal{F}}_{5} & = & -A_{3} + A_{4} ~, \\
{\cal{F}}_{6} & = & \frac{1}{E_{N} + m_{N}}~\Bigl[~\{2P \cdot 
k\,(E_{N} - E_{Y})+({\textstyle \frac{1}{2}} k^{2} - k \cdot q_K)
(E_{N} + E_{Y})+ P \cdot kk_{0}\}A_{2} \nonumber\\
 && \hspace{2.2cm} + ( k_0 E_K - k \cdot q_K ) A_{3} + 
\{k_{0}(E_{N} + E_{Y}) - 2P \cdot k \}A_{4}  
\nonumber\\
 & & \hspace{2.2cm} - ( k_0 k \cdot q_K - 
k^2 E_K ) A_{5} 
+ (k^2 - k_0^2) A_{6}~\Bigr] ~, \\
{\cal{F}}_{7} & = & \frac{1}{E_{N} + m_{N}}~\Bigl[\,A_{1} - P \cdot k A_{2} 
- k_{0}A_{3} - (m_{N} + m_{Y})A_{4} - (k^{2} - k \cdot q_K)A_{5} 
 + k_{0}A_{6}\,\Bigr] , \\
{\cal{F}}_{8} & = & \frac{1}{E_{N} + m_{N}}~\Bigl[-(2P \cdot k + 
{\textstyle \frac{1}{2}} k^{2} - k \cdot q_K)A_{2} - k_{0}(A_{3} + A_{4}) - 
k^{2}A_{5}~\Bigr] ~, \\
{\cal{F}}_{9} & = & \frac{1}{E_{N}+m_{N}}~\Bigl[~(2P \cdot k - 
{\textstyle \frac{1}{2}} k^{2} + k \cdot q_K)A_{2} + k_{0}(A_{3} - A_{4}) + 
k^{2}A_{5}~\Bigr] ~, \\
{\cal{F}}_{10} & = & \frac{1}{E_{Y} + m_{Y}}~\Bigl[-\{2P \cdot 
k(E_{N} - E_{Y}) + ({\textstyle \frac{1}{2}} k^{2} - k \cdot q_K)
(E_{N} + E_{Y}) + P \cdot k k_{0} \}A_{2}\nonumber\\
 & & \hspace{2.1cm}  + (k_0E_K - k \cdot q_K) A_{3} + 
\{k_{0}(E_{N} + E_{Y}) - 2P \cdot k \}A_{4}  \nonumber\\
 & & \hspace{2.1cm}  + ( k_0 k \cdot q_K - 
k^2 E_K ) A_{5} + (k^2-k_0^2) A_{6}~\Bigr] ~, \\
{\cal{F}}_{11} & = & \frac{1}{E_{Y} + m_{Y}}~\Bigl[-A_{1}+
P\cdot kA_{2} - k_{0}A_{3} + (m_{N} + m_{Y})A_{4} + (k^{2} - 
k \cdot q_K)A_{5} + k_{0}A_{6}\,\Bigr] ,\nonumber\\ \\
{\cal{F}}_{12} & = & \frac{1}{E_{Y} + m_{Y}}~\Bigl[~(2P \cdot k + 
{\textstyle \frac{1}{2}} k^{2} - k \cdot q_K)A_{2} - k_{0}(A_{3} + A_{4}) + 
k^{2}A_{5}~\Bigr] ~, \\
{\cal{F}}_{13} & = & \frac{1}{E_{Y} + m_{Y}}~\Bigl[-(2P \cdot k 
+ k \cdot q_K - {\textstyle \frac{1}{2}} k^{2})A_{2} + k_{0}(A_{3} - A_{4}) 
- k^{2}A_{5}~\Bigr] ~, \\
{\cal{F}}_{14} & = & \frac{1}{E_{N} + m_{N}}~\Bigl[-A_{1}
 + (m_{N} + m_{Y})A_{4}~\Bigr] ~, \\
{\cal{F}}_{15} & = & \frac{1}{E_{Y} + m_{Y}}~\Bigl[~A_{1}
 - (m_{N} + m_{Y})A_{4}~\Bigr] ~, \\
{\cal{F}}_{16} & = & \frac{1}{(E_{N} + m_{N})(E_{Y} + m_{Y})} 
\Bigl[-k_{0}A_{1} + k \cdot q_K A_{3} + \{2P \cdot k + k_{0}(m_{N} + 
m_{Y})\}A_{4} - k^{2}A_{6}\Bigr] ,
\nonumber\\
\end{eqnarray}

\begin{eqnarray}
{\cal{F}}_{17} & = & \frac{1}{(E_{N} + m_{N})(E_{Y} + m_{Y})}\,
\Bigl[\,A_{1} - E_K A_{3} - (E_{N} + E_{Y}  
+ m_{N} + m_{Y})A_{4} + k_{0}A_{6}\,\Bigr] ~,
\nonumber\\ \\
{\cal{F}}_{18} & = & \frac{1}{(E_{N} + m_{N})(E_{Y} + m_{Y})} 
~\Bigl[~A_{3} - A_{6}~\Bigr] ~, \\
{\cal{F}}_{19} & = & \frac{1}{(E_{N} + m_{N})(E_{Y} + m_{Y})} 
~\Bigl[~A_{3} + A_{4}~\Bigr] ~, \\
{\cal{F}}_{20} & = & \frac{1}{(E_{N} + m_{N})(E_{Y} + m_{Y})} 
~\Bigl[-A_{3} + A_{4}~\Bigr] ~.
\end{eqnarray}

\section{Normalization of the Three-Body Wave Functions}
\label{derivation_sqrt3}
In the $\gamma + {^3{\rm He}} \to \pi^+ + {^3{\rm H}}$ the initial
and final nuclear wave functions are anti-symmetrized. Therefore,
in this case 
it is sufficient to evaluate the elementary production on one of
the nucleons and the nuclear amplitude is multiplied with an 
antisymmetry factor 3~\cite{tiator2}. In the hypertriton
productions [see Eqs.~(\ref{eq:hyp_photo}) and (\ref{eq:hyp_electro})],
the final nucleus consists of two nucleons and one hyperon.
As a consequence, a proper normalization is required, and
the anti-symmetry factor has to be recalculated. To this end,
we will make use of 
the method of second quantization, i.e.,
\begin{eqnarray}
\left\{ N^{\dagger}(x),N(y) \right\} &=& \delta_{x,y} ~, \\
\left\{ N^{\dagger}(x),\Lambda (y) \right\} &=& 0 ~,
\end{eqnarray}
we can show that the normalized $^3$He and hypertriton wave functions
can be written as
\begin{eqnarray}
\left|~ ^3{\rm He}~ \right\rangle &=& \frac{1}{\sqrt{3!}}~ \int dx dy dz
~ N^{\dagger}(x) N^{\dagger}(y) N^{\dagger}(z)~ \left| ~0~ \right\rangle ~
\Phi_{NNN}(x,y,z) ~, \\
\left|~ ^3_\Lambda{\rm H}~ \right\rangle &=& \frac{1}{\sqrt{2!}}~ \int
dx dy dz ~ N^{\dagger}(x) N^{\dagger}(y) \Lambda^{\dagger}(z) ~ \left|
~0~ \right\rangle ~ \Phi_{NN\Lambda}(x,y,z) ~,
\end{eqnarray}
where $N^{\dagger}(x)$ [$\Lambda^{\dagger}(z)$] represents the nucleon 
[$\Lambda$] creation operator at the point $x$ [$z$], while $\Phi_{NNN}$ 
and $\Phi_{NN\Lambda}$ denote the spatial $^3$He and hypertriton
wave functions, respectively.

To create a $\Lambda$ hyperon from a nucleon we need the one-body operator
\begin{eqnarray} 
{\cal O}(x) &=& \int dx~ \Lambda^{\dagger}(x)~ h(x)~ N(x) ~, 
\end{eqnarray} 
where $h(x)$ indicates the elementary operator, which
can be sandwiched between the $^3$He and the hypertriton wave functions to
give 
\begin{eqnarray} 
&&\left\langle ~^3_\Lambda{\rm H} ~\right|~ {\cal O}(x)~
\left| ~^3{\rm He}  ~\right\rangle ~=~ \frac{1}{\sqrt{12}} \int dx dy dz
 \nonumber\\
&& \times ~\Bigl\{ \Phi^*_{NN\Lambda}(y,z,x) ~h(x)~ \Phi_{NNN}(x,y,z)
 - \Phi^*_{NN\Lambda}(z,y,x) ~h(x)~ \Phi_{NNN}(x,y,z) \nonumber \\
&& - ~\Phi^*_{NN\Lambda}(x,z,y) ~h(y)~ \Phi_{NNN}(x,y,z) 
 + \Phi^*_{NN\Lambda}(z,x,y) ~h(y)~ \Phi_{NNN}(x,y,z) \nonumber\\
&& + ~\Phi^*_{NN\Lambda}(x,y,z) ~h(z)~ \Phi_{NNN}(x,y,z) 
 - \Phi^*_{NN\Lambda}(y,x,z) ~h(z)~ \Phi_{NNN}(x,y,z) \Bigr\} ~. 
\label{allxyz} 
\end{eqnarray} 
Note that every term in Eq.~(\ref{allxyz}) is integrated over $x$, $y$, and 
$z$. By making use of the anti-symmetric behavior of $\Phi_{NNN}(x,y,z)$,
we can recast Eq.~(\ref{allxyz}) to 
\begin{eqnarray} 
&&\left\langle ~^3_\Lambda{\rm H} ~\right|~ {\cal O}(x)~
\left| ~^3{\rm He}  ~\right\rangle ~=~ \frac{1}{\sqrt{12}} \int dx dy dz
 \nonumber\\
&& \times ~\Bigl\{ \Phi^*_{NN\Lambda}(y,z,x) ~h(x)~ \Phi_{NNN}(y,z,x)
 + \Phi^*_{NN\Lambda}(z,y,x) ~h(x)~ \Phi_{NNN}(z,y,x) \nonumber \\
&& + ~\Phi^*_{NN\Lambda}(x,z,y) ~h(y)~ \Phi_{NNN}(x,z,y) 
 + \Phi^*_{NN\Lambda}(z,x,y) ~h(y)~ \Phi_{NNN}(z,x,y) \nonumber\\
&& + ~\Phi^*_{NN\Lambda}(x,y,z) ~h(z)~ \Phi_{NNN}(x,y,z) 
 + \Phi^*_{NN\Lambda}(y,x,z) ~h(z)~ \Phi_{NNN}(y,x,z) \Bigr\} \nonumber\\
&=& \sqrt{3} \int dx dy dz ~
\Phi^*_{NN\Lambda}(x,y,z) ~h(z)~ \Phi_{NNN}(x,y,z) ~,
\end{eqnarray}
which shows that the required anti-symmetry factor for the hypertriton
production off $^3$He is $\sqrt{3}$.
If we used the above prescription to calculate the amplitude of pion
photoproduction on $^3$He, where the one-body operator in this case 
may be written as 
\begin{eqnarray} 
{\cal O}(x) &=& \int dx~ N^{\dagger}(x)~ h(x)~ N(x) ~, 
\end{eqnarray} 
we would obtain
\begin{eqnarray} 
\left\langle ~{^3{\rm H}} ~\right|~ {\cal O}(x) ~\left| ~^3{\rm He}  
~\right\rangle &=& 3 \int dx dy dz ~
\Phi^*_{NNN}(x,y,z) ~h(z)~ \Phi_{NNN}(x,y,z) ~,
\end{eqnarray}
which is consistent with previous work \cite{tiator2}.

\section{Components of the Elementary Operator Matrix}
\label{component_of_jmu}
Here we give the individual components of the matrix [j], defined
in Eq.~(\ref{eq:define_j}) through the relation
 $J^\mu=[\sigma]\,[j]$, which are useful for the numerical calculation
of the observables.
\begin{eqnarray}
  j_{00} &=& iN{\cal{F}}_{17}\,\bvec{p}_{N} 
    \cdot (\bvec{p}_Y \times \bvec{k})~,\\
  j_{\ell 0} &=& iN \Bigl[{\cal F}_{14}\, \bvec{p}_N \times \bvec{k}+
{\cal F}_{15}\,\bvec{p}_Y \times \bvec{k} -{\cal F}_{16}\,\bvec{p}_N
 \times \bvec{p}_Y - 
\bvec{p}_N\cdot\bvec{p}_Y \times \bvec{k} 
\left({\cal F}_{18}\, \bvec{k}\, + \right.\nonumber\\ && \left.
{\cal F}_{19}\, \bvec{p}_N + {\cal F}_{20}\, \bvec{p}_Y\right)\Bigr]_l~, \\
  j_{0m} &=& N \{({\cal F}_2- \bvec{p}_{N} 
    \cdot \bvec{p}_Y\,{\cal F}_{17})\,k_m+
        ({\cal F}_6+ \bvec{p}_{Y} 
    \cdot \bvec{k}\,{\cal F}_{17})\,p_{N,m}+
        ({\cal F}_{10}+ \bvec{p}_{N} 
    \cdot \bvec{k}\,{\cal F}_{17})\,p_{Y,m}
        \} ,~~~~\\
  j_{\ell m} &=& A\,\delta_{\ell m} + {B}_{\ell}\,k_m + 
  {C}_\ell\,p_{N,m}  + {D}_\ell\,p_{Y,m}  ~,
\end{eqnarray}
where $\ell,m=x,y,z$, and 
\begin{eqnarray} 
A &=& -N \left[{\cal F}_1+{\cal F}_{14}\,\bvec{p}_N\cdot\bvec{k} -
{\cal F}_{15}\,\bvec{p}_Y\cdot\bvec{k}-{\cal F}_{16}\,\bvec{p}_N
\cdot\bvec{p}_Y\right]
~,\\
\bvec{B} &=& -N \left[({\cal{F}}_{3}-
  \bvec{p}_{N}\cdot\bvec{p}_Y\; {\cal{F}}_{18})\; \bvec{k}
  + ( {\cal{F}}_{4}-{\cal{F}}_{14}-\bvec{p}_{N}\cdot\bvec{p}_Y\;
  {\cal{F}}_{19})\;\bvec{p}_{N}\right. \nonumber\\
  && \hspace{22mm}\left.
  +~ ( {\cal{F}}_{5}+{\cal{F}}_{15}-\bvec{p}_{N}\cdot\bvec{p}_Y\;
  {\cal{F}}_{20})\;\bvec{p}_{Y}
\right] ~,\\
\bvec{C} &=& -N \left[ ({\cal{F}}_{7}+{\cal F}_{14} + 
  \bvec{p}_{Y}\cdot\bvec{k}\; {\cal{F}}_{18})\; \bvec{k}
  + ( {\cal{F}}_{8}+\bvec{p}_{Y}\cdot\bvec{k}\;
  {\cal{F}}_{19})\;\bvec{p}_{N}\right. \nonumber\\
  && \hspace{22mm}\left.
  +~ ( {\cal{F}}_{9}+{\cal{F}}_{16}+\bvec{p}_{Y}\cdot\bvec{k}\;
  {\cal{F}}_{20})\;\bvec{p}_{Y}
 \right] ~,
\\
\bvec{D} &=& -N \left[ ({\cal{F}}_{11}+{\cal F}_{15} + 
  \bvec{p}_{N}\cdot\bvec{k}\; {\cal{F}}_{18})\; \bvec{k}
  +~ ( {\cal{F}}_{12}+{\cal F}_{16}+\bvec{p}_{N}\cdot\bvec{k}\;
  {\cal{F}}_{19})\;\bvec{p}_{N}\right. \nonumber\\
  && \hspace{22mm}\left.
  +~ ( {\cal{F}}_{13}+\bvec{p}_{N}\cdot\bvec{k}\;
  {\cal{F}}_{20})\;\bvec{p}_{Y}
 \right]
~.
\end{eqnarray}

\section{Useful Kinematical Relations in the Nuclear System}
In the $\gamma-^3$He laboratory system the energy and momentum of 
the virtual photon are obtained from 
\begin{eqnarray}
  k_0 &=& \left(W^2-M_{\rm He}^2-k^2\right)/(2M_{\rm He})~~~,~~~
  |\bvec{k}| ~=~ (k_0^2-k^2)^{1/2}~,
\end{eqnarray}
where we denote the total c.m. energy by $W$.
On the other hand the energy and momentum of the kaon in the c.m.
frame are given by
\begin{eqnarray}
  E_K^{\rm c.m.} &=& \left(W^2-m_\Lambda^2+m_K^2\right)/(2W)~~~,~~~
  |\bvec{q}_K^{\rm c.m.}| ~=~ \{(E_K^{\rm c.m.})^2-m_K^2\}^{1/2}~.
\end{eqnarray}
The corresponding energy and momentum in the laboratory frame are
obtained from the following transformation
\begin{eqnarray}
  E_K &=& \gamma\Bigl(E_K^{\rm c.m.}+v|\bvec{q}_K^{\rm c.m.}|\cos\theta_{\rm c.m.}\Bigr)
  ~~~,~~~ |\bvec{q}_K| ~=~ (E_K^2-m_K^2)^{1/2}~,
\end{eqnarray}
whereas the kaon scattering angle in the laboratory frame is given by
\begin{eqnarray}
  \cos\theta &=& \gamma\Bigl(|\bvec{q}_K^{\rm c.m.}|\cos\theta_{\rm c.m.}+vE_K^{\rm c.m.}\Bigr)~,
\end{eqnarray}
where
\begin{eqnarray}
  \gamma &=& (k_0+M_{\rm He})/W ~~~,~~~ v ~=~ k/(k_0+M_{\rm He})~.
\end{eqnarray}

\begin{figure}[!t]
  \begin{center}
    \leavevmode
    \epsfig{figure=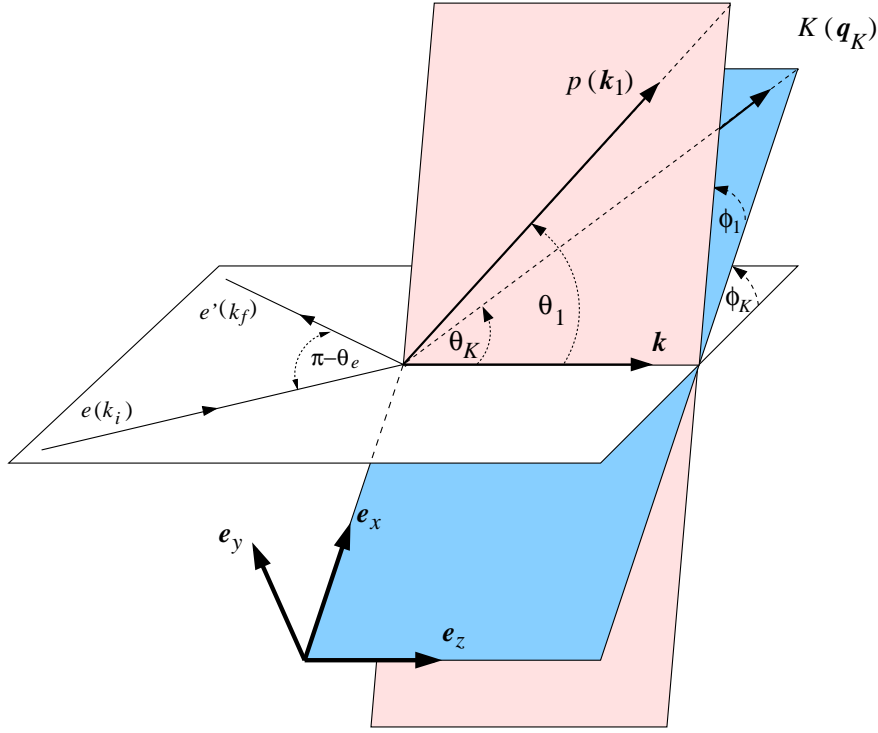,width=120mm}
    \caption{(Color online) Kinematics of the hypertriton
        electroproduction on $^3$He. Note that the $x-z$ plane
        is defined as the $\gamma_v-K$ production plane.}
   \label{fig:kinematics} 
  \end{center}
\end{figure}

The formulas below are derived according to Fig.~\ref{fig:kinematics} in the 
laboratory system with Jacobi coordinates. 
These relations have been used in the numerical calculations.
\label{kinematics}
We choose the kaon scattering plane as the $xz$-plane, whereas
the direction of virtual photon three-momentum defines
the $z$-direction, i.e.,
\begin{eqnarray}
\bvec{k} &=& k\, \bvec{e}_{z} ~.  
\end{eqnarray}
Consequently, 
the initial momentum of the first nucleon and the momentum of the 
kaon are given by
\begin{eqnarray}
\bvec{k}_1 &=& k_1~(\sin \theta_{1} \cos \phi_{1} \bvec{e}_{x} + \sin 
\theta_{1} 
\sin \phi_{1} \bvec{e}_{y} + \cos \theta_{1} \bvec{e}_{z})~ ,\\
\bvec{q}_K &=& q_K~(\sin \theta_{K} 
\bvec{e}_{x} + \cos \theta_{K} \bvec{e}_{z}) ~,
\end{eqnarray}
while the momentum of the produced hyperon is given by 
\begin{eqnarray}
\bvec{k}_1' &=& \bvec{k}_1+{\textstyle \frac{2}{3}}\left( \bvec{k}-\bvec{q_K}\right)\nonumber\\
        &=& (k_1\sin\theta_1\cos\phi_1-{\textstyle \frac{2}{3}}q_K\sin\theta_K)\,\bvec{e}_x+(k_1\sin\theta_1\sin\phi_1)\,\bvec{e}_y\nonumber\\
        &&+(k_1\cos\theta_1+{\textstyle \frac{2}{3}}k-{\textstyle \frac{2}{3}}q_K\cos\theta_K)\,\bvec{e}_z\nonumber\\
        &\equiv& k_1'\left(\sin\theta_1'\cos\phi_1'\,\bvec{e}_x +
        \sin\theta_1'\sin\phi_1'\,\bvec{e}_y +\cos\theta_1'\,\bvec{e}_z\right) ~.
\end{eqnarray}
From this we can derive the following expressions for 
vector and scalar products of the photon ($\bvec{k}$), 
nucleon ($\bvec{k}_1$), kaon ($\bvec{q}_K$),
and hyperon ($\bvec{k}_1'$) momenta,
\begin{eqnarray}
\bvec{q}_K \cdot \bvec{k} &=& q_K k~\cos\theta_{K} ~, \\
\bvec{k}_1 \cdot \bvec{k} &=& k_1 k~\cos\theta_{1} ~, \\
\bvec{q}_K \cdot \bvec{k}_1 &=& q_K k_1~(\sin\theta_{K}~\sin\theta_{1}~\cos 
\phi_{1} + \cos\theta_{K}~\cos\theta_{1}) ~, \\
\bvec{k}_1 \cdot (\bvec{k} - \bvec{q}_K)
&=& k_1k~\cos\theta_{1} - k_1q_K~(\sin\theta_{K}~\sin\theta_{1}~\cos\phi_{1}
+ \cos\theta_{K}~\cos\theta_{1}) ~, \\
\bvec{q}_K \times \bvec{k}_1 &=& q_Kk_1~[-\sin \theta_{1} \sin 
\phi_{1}~\cos \theta_{K}~\bvec{e}_{x} + (\sin \theta_{1} \cos 
\phi_{1}~\cos \theta_{K}\nonumber\\ 
&&- \sin\theta_{K}~\cos\theta_{1})~\bvec{e}_{y}
+ \sin\theta_{1}~\sin\phi_{1}~\sin\theta_{K}~\bvec{e}_{z}] ~, \\
\bvec{k} \times \bvec{q}_K &=& kq_K~\sin\theta_{K}~\bvec{e}_{y} ~, \\
\bvec{k}_1 \times \bvec{k} &=& k_1 k~\sin\theta_{1}~(\sin \phi_{1} 
\bvec{e}_{x} - \cos \phi_{1} \bvec{e}_{y}) ~, \\
\bvec{k}_1' \times \bvec{k}_1 &=& {\textstyle \frac{2}{3}}~(\bvec{k} \times 
\bvec{k}_1 - \bvec{q}_K \times \bvec{k}_1) ~,\\
k_1' &=& \left\{{k}_1^2 +{\textstyle \frac{4}{3}}(\bvec{k}_1\cdot\bvec{k}
        -\bvec{k}_1\cdot\bvec{q}_K)+{\textstyle \frac{4}{9}}
        ({k}^2+{q}_K^2-2\bvec{k}\cdot\bvec{q}_K)
        \right\}^{1/2}~,\\
\theta_1' &=& \cos^{-1}\Bigl[\left\{
        k_1\cos\theta_1+{\textstyle \frac{2}{3}}(k-
        q_K\cos\theta_K)\right\}/k_1'\Bigr] ~,\\
\phi_1' &=& \sin^{-1}\{(k_1\sin\theta_1\sin\phi_1)/(k_1'\sin\theta_1')\}~.
\end{eqnarray}


\begin{thebibliography}{99}
\bibitem{sighyp} K. Miyagawa, private communication. 
\bibitem{Afnan}  I. R. Afnan
                 and B. F. Gibson, Phys. Rev. C. {\bf 47}, 1000 (1993).
\bibitem{Dover} C. B. Dover, H. Feshbach, and A. Gal, Phys. Rev. C {\bf 51}, 
                 541 (1995).
\bibitem{Juric:1973zq} M.~Juric {\it et al.}, Nucl.\ Phys.\  B {\bf 52}, 1 (1973).
\bibitem{oldi} G. Keyes {\it et al}., Phys. Rev. Lett. {\bf 20}, 819 (1968);
               {\it ibid.}, Phys. Rev. D {\bf 1}, 66 (1970);
               G. Keyes, J. Sacton, J. H.
               Wickens, and M. M. Block, Nucl. Phys. {\bf B67}, 269 (1973);
               R. E. Phillips and J. Schneps, Phys. Rev. Lett. {\bf 20},
               1383 (1968); {\it ibid.}, Phys. Rev. {\bf 180}, 1307 (1969);
               G. Bohm {\it et al}., Nucl. Phys. {\bf B16}, 46 (1970).
\bibitem{miyagawa93} K. Miyagawa and W. Gl\"ockle, Phys. Rev. C {\bf 48},
                   2576 (1993).
\bibitem{miyagawa95} K. Miyagawa, H. Kamada, W. Gl\"ockle, and V. Stoks,
                  Phys. Rev. C {\bf 51}, 2905 (1995).
\bibitem{hyp_properties} J.~Golak {\it et al.},
        Phys.\ Rev.\  C {\bf 55}, 2196 (1997);
        A.~Cobis, A.~S.~Jensen and D.~V.~Fedorov,
        J.\ Phys.\ G {\bf 23}, 401 (1997);
        H.~Kamada, J.~Golak, K.~Miyagawa, H.~Witala and W.~Gloeckle,
        Phys.\ Rev.\  C {\bf 57}, 1595 (1998);
        J.~Golak, H.~Witala, K.~Miyagawa, H.~Kamada and W.~Gloeckle,
        Phys.\ Rev.\ Lett.\  {\bf 83}, 3142 (1999);
        K.~Tominaga and T.~Ueda,
        Nucl.\ Phys.\ A {\bf 693}, 731 (2001);
        H.~W.~Hammer, Nucl.\ Phys.\  A {\bf 705}, 173 (2002);
        H.~Nemura, Y.~Akaishi and Y.~Suzuki,
        Phys.\ Rev.\ Lett.\  {\bf 89}, 142504 (2002);
        D.~V.~Fedorov and A.~S.~Jensen,
        Nucl.\ Phys.\ A {\bf 697}, 783 (2002);
        Y.~Fujiwara, K.~Miyagawa, M.~Kohno and Y.~Suzuki,
        Phys.\ Rev.\  C {\bf 70}, 024001 (2004).
\bibitem{juelich} A. G. Reuber, K. Holinde, and J. Speth, Czech. 
        J. Phys. {\bf 42}, 1115 (1992); K. Holinde, Nucl. Phys. 
        A {\bf 547}, 255c (1992).
\bibitem{nijmegen89} P. M. M. Maessen, Th. A. Rijken, and J. J. de Swart,
                    Phys. Rev. C 40 (1989) 2226.
\bibitem{komarov} V. I. Komarov, A. V. Lado, and Yu. N. Uzikov, J. Phys. 
        G {\bf 21}, L69 (1995).
\bibitem{Gardestig:1996zz} A.~Gardestig, Z.\ Phys.\  A {\bf 357}, 101 (1997).
\bibitem{mart98} T. Mart, L. Tiator, D. Drechsel, and C. Bennhold,
                 Nucl. Phys. {\bf A640}, 235 (1998).
\bibitem{mart_thesis} T. Mart, Ph.D. Thesis, Universit\"at Mainz, 1996 (unpublished).
\bibitem{kim} R. A. Brandenburg, Y. E. Kim, and A. Tubis, Phys. Rev. C
              {\bf 12}, 1368  (1975).
\bibitem{congleton} J. G. Congleton, J. Phys. G {\bf 18}, 339 (1992).
\bibitem{williams} R. A. Williams, C.-R. Ji, and S. R. Cotanch, Phys. Rev. 
                    D {\bf 41}, 1449 (1990); Phys. Rev. C {\bf 43}, 452
                    (1991); Phys. Rev. C {\bf 46}, 1617 (1992).
\bibitem{Dohrmann:2004xy}
                  F.~Dohrmann {\it et al.},
                  Phys.\ Rev.\ Lett.\  {\bf 93}, 242501 (2004).
\bibitem{note_on_dohrmann} See the upper panel of Fig.\,3 of Ref.~\cite{Dohrmann:2004xy}.
\bibitem{note_on_dohrmann1} See the lower panel of Fig.\,3 of Ref.~\cite{Dohrmann:2004xy}.
\bibitem{nijmegen93} V. G. J. Stoks, R. A. M. Klomp, C. P. F. Terheggen,
                     and J. J. de Swart, Phys. Rev. C {\bf 49}, 2950 (1994).
\bibitem{kaon-maid}T. Mart and C. Bennhold, Phys. Rev. C {\bf 61}, 012201(R) (1999);
               T.~Mart, Phys.\ Rev.\ C {\bf 62}, 038201 (2000); C.~Bennhold,
               H.~Haberzettl and T.~Mart, arXiv:nucl-th/9909022;
               T. Mart, C. Bennhold, H. Haberzettl, and L. Tiator,
               http://www.kph.uni-mainz.de/MAID/kaon/kaonmaid.html.
\bibitem{YaM99} H. Yamamura, K. Miyagawa, T. Mart, C. Bennhold,
        W. Gl\"ockle, and H. Haberzettl, 
        Phys. Rev. C {\bf 61}, 014001 (1999);
        K.~Miyagawa, T.~Mart, C.~Bennhold and W.~Gl\"ockle,
        Phys. Rev. C {\bf 74}, 034002 (2006);
        A.~Salam, K.~Miyagawa, T.~Mart, C.~Bennhold and W.~Gl\"ockle,
        Phys. Rev. C {\bf 74}, 044004 (2006).
\bibitem{Haberzettl:1998eq} H.~Haberzettl, C.~Bennhold, T.~Mart, and 
        T.~Feuster, Phys. Rev. C {\bf 58}, R40 (1998).
\bibitem{saphir} 
   M. Q. Tran {\it et al.}, Phys. Lett. B {\bf 445}, 20 (1998).
\bibitem{niculescu} G. Niculescu {\it et al.}, {\it Phys. Rev. Lett.} {\bf 81},
              1805 (1998).
\bibitem{mohring} R. M. Mohring {\it et al.}, Phys. Rev. C {\bf 67}, 055205 (2003).
\bibitem{old_electro} P. Brauel {\it et al}., {\it Z. Phys.} C {\bf 3}, 101 (1979).
\bibitem{photon_point} A. Bleckman {\it et al}., Z. Phys. {\bf 239}, 1 (1970).
\bibitem{Glander:2003jw}
  K.~H.~Glander {\it et al.},
  Eur.\ Phys.\ J.\ A {\bf 19}, 251 (2004).
\bibitem{Tran:1998qw}
  M.~Q.~Tran {\it et al.}, Phys.\ Lett.\ B {\bf 445}, 20 (1998).
\bibitem{Bradford:2005pt}
  R.~Bradford {\it et al.},
  Phys.\ Rev.\ C {\bf 73}, 035202 (2006).
\bibitem{old_data} References for older data are given, e.g., in
        Ref.~\cite{saphir}.
\bibitem{Mart:2006dk}
  T.~Mart and A.~Sulaksono,
  Phys.\ Rev.\ C {\bf 74}, 055203 (2006).
\bibitem{Bydzovsky:2006wy} See the Introduction part of 
  P.~Bydzovsky and T.~Mart,
  Phys.\ Rev.\  C {\bf 76}, 065202 (2007); and
  P.~Bydzovsky, M.~Sotona, T.~Motoba, K.~Itonaga, K.~Ogawa and O.~Hashimoto,
  arXiv:0706.3836 [nucl-th].
\bibitem{deshalit63} A. de Shalit and I. Thalmi, {\it Nuclear Shell Theory}
                   (Academic Press, New York, 1963).
\bibitem{tiator81} L. Tiator and D. Drechsel, Nucl. Phys. A {\bf 360},
                   208 (1981).
\bibitem{chew} G. F. Chew, M. L. Goldberger, F. E. Low, and Y. Nambu,
               Phys. Rev. {\bf 106}, 1345 (1957).
\bibitem{tiator2} L. Tiator, A. K. Rej, and D. Drechsel, Nucl. Phys.
                  A {\bf 333}, 343 (1980); L. Tiator, Nucl. Phys.
                  A {\bf 364}, 189 (1981).
\bibitem{tia_th} L. Tiator, Ph. D. Thesis, Universit\"at Mainz, 1980 (unpublished).
\bibitem{Adelseck:1986fb} R.~A.~Adelseck, C.~Bennhold and L.~E.~Wright,
  Phys.\ Rev.\  C {\bf 32}, 1681 (1985).
\end{thebibliography}
\end{document}